\documentclass[fleqn,usenatbib]{mnras}

\usepackage[T1,T4]{fontenc}
\usepackage{ae,aecompl}

\DeclareRobustCommand{\VAN}[3]{#2}
\let\VANthebibliography\thebibliography
\def\thebibliography{\DeclareRobustCommand{\VAN}[3]{##3}\VANthebibliography}

\usepackage{graphicx}	
\usepackage{amsmath}	
\usepackage{txfonts}
\usepackage[usenames,dvipsnames]{xcolor}
\usepackage{url}
\usepackage{array}
\usepackage{longtable}
\usepackage{float}
\usepackage{pdflscape}
\usepackage{soul}
\usepackage[normalem]{ulem}

\title[\texttt{SIMBA}-\texttt{C} chemical enrichment of the ISM]{\texttt{SIMBA}-\texttt{C}: An updated chemical enrichment model for galactic chemical evolution in the \texttt{SIMBA} simulation}

\author[R. T. Hough et al.]{
Renier T. Hough,$^{1}$\thanks{E-mail: renierht@gmail.com}
Douglas Rennehan,$^{2}$
Chiaki Kobayashi,$^{3}$
S. Ilani Loubser,$^{1, 4}$
Romeel Dav\'e,$^{5,6}$
\newauthor Arif Babul,$^{7}$ and
Weiguang Cui$^{4,8,9}$
\\
$^{1}$Centre for Space Research, North-West University, Potchefstroom 2520, South Africa\\
$^{2}$Center for Computational Astrophysics, Flatiron Institute, 162 5th Ave, New York, NY 10010, USA\\
$^{3}$Centre for Astrophysics Research, Department of Physics, Astronomy and Mathematics, University of Hertfordshire, Hatfield AL10 9AB, UK\\
$^{4}$National Institute for Theoretical and Computational Sciences (NITheCS), Potchefstroom 2520, South Africa\\
$^{5}$School of Physics and Astronomy, University of Edinburgh, Edinburgh EH9 3HJ, UK\\
$^{6}$Department of Physics and Astronomy, University of the Western Cape, Bellville, 7535, South Africa\\
$^{7}$Department of Physics \& Astronomy, University of Victoria, Victoria BC V8P 5C2, Canada\\
$^{8}$Departamento de Física Teórica, M-8, Universidad Autónoma de Madrid, Cantoblanco 28049, Madrid, Spain\\
$^{9}$Centro de Investigación Avanzada en Física Fundamental (CIAFF), Universidad Aut\'{o}noma de Madrid, Cantoblanco, 28049 Madrid, Spain
}

\date{Accepted XXX. Received YYY; in original form ZZZ}

\pubyear{2023}

\begin{document}
\label{firstpage}
\pagerange{\pageref{firstpage}--\pageref{lastpage}}

\maketitle

\begin{abstract}
We introduce a new chemical enrichment and stellar feedback model into \texttt{GIZMO}, using the \texttt{SIMBA} sub-grid models as a base. Based on the state-of-the-art chemical evolution model of Kobayashi et al., \texttt{SIMBA-C} tracks 34 elements from H$\rightarrow$Ge and removes \texttt{SIMBA}'s instantaneous recycling approximation. Furthermore, we make some minor improvements to \texttt{SIMBA}'s base feedback models.  \texttt{SIMBA-C} provides significant improvements on key diagnostics such as the knee of the $z=0$ galaxy stellar mass function, the faint end of the main sequence, and the ability to track black holes in dwarf galaxies.  \texttt{SIMBA-C} also matches better with recent observations of the mass-metallicity relation at $z=0,2$.  By not assuming instantaneous recycling, \texttt{SIMBA-C} provides a much better match to galactic abundance ratio measures such as [O/Fe] and [N/O].  \texttt{SIMBA-C} thus opens up new avenues to constrain feedback models using detailed chemical abundance measures across cosmic time.
\end{abstract}




\section{Introduction}
Studying the large-scale structure of the Universe, specifically the formation of galaxies, groups of galaxies, and clusters of galaxies and how they evolve over time, can give us insight into the inner workings of how the Universe formed and evolved \citep{Dolag2008}. Numerical simulations can help to visualise and understand the key physical processes involved in galaxy and other large-scale structure evolution\footnote{For more insight into these simulations, we refer the reader to \citep{Bertschinger1998, Springel2006, Dolag2008, Dave2011, Hahn2011, Somerville2015, Naab2017, Vogelsberger2020, Hough2021, Oppenheimer2021, Jung2022, Kobayashi2023}.}. One such process is the rate at which stars form. 

It is well documented that free-falling cool gas leads to the formation of stars \citep{Kennicutt2007, Leroy2008}. However, this alone cannot account for the observed rate at which stars are formed \citep{Rasia2015}. We know that cold metal-free gas ($T<10^{4}\, \mathrm{ K}$) cooling solely through rotational-vibrational lines of H$_{2}$ has a minimum achievable temperature of $\sim 200\, \mathrm{ K}$ \citep{Smith2009}. The energy levels of H$_{2}$ become thermalised at low densities ($n\sim 10^{3}-10^{4} \,\mathrm{cm} ^{-3}$), above which the cooling rate is $\propto n$ \citep{Bromm2002, Smith2009}. This leads to a stalling point in the star formation process as the gas cooling time increases above the dynamical time \citep{Abel2002, Stinson2013}. 

Other processes must therefore be involved. The first process is metal-cooling, where metals are created in stars through the nucleosynthesis processes, due to the early star formation from metal-free cold gas. These metals then provide more available atomic and molecular transitions, allowing the gas to lose its internal energy faster than in the metal-free case \citep{Smith2008}. This reduces the gas cooling time to be shorter than the dynamical time and star formation can proceed again. Therefore, metals within cold collapsing gas clouds increase the efficiency of star formation \citep{Kobayashi2007, Smith2009}. However, this will result in a star formation rate (SFR) that can be higher than observed \citep{Stinson2013, Rasia2015}, in turn leading to too many old spheroidal galaxies at $z=0$ \citep{White1991, Piontek2011}. 

Therefore processes, called feedback, e.g. supernovae (SNe) and Active Galactic Nuclei (AGN), are needed to re-heat the gas \citep{Piontek2011,Stinson2013}. These processes inject massive amounts of energy into the cold collapsing gas clouds, which increases the temperature and stalls star formation. This quenching of star formation can be seen in the famous SFR vs cosmic time result from \citet{Madau2014} after the SFR reached a peak at $z\sim 2$, called Cosmic High Noon \citep{Mehta2017}. However, events such as stellar winds and SNe inject new types of metals, leading to a self-regulating cycle between star formation and feedback\footnote{The interaction of these processes has been extensively studied using observations (\citealt{Ceverino2009, Smith2009, Piontek2011, Hopkins2012, Hirschmann2016, Romano2019, Lagos2022} and references therein).}.

Studying the chemical composition of these newly created and injected metals within galaxies can provide us with powerful insights into all of the factors discussed above, i.e. star formation efficiency, gas inflow rate, and metal-rich gas outflow rate \citep{Maiolino2019, Beverage2023}. Furthermore, the composition of the stellar content of galaxies contains the integrated enrichment of the gas over its entire star formation history \citep{Cameron1971}. This is particularly true for stellar abundances in quiescent galaxies, which have prominent absorption features.

Two of the well-known properties of massive quiescent galaxy properties are: i) The most massive local quiescent galaxies contain the most metals \citep{Kobayashi1999, Gallazzi2005, Thomas2005, Beverage2023}. This is thought to be a reflection of the strength of the gravitational potential of the host galaxy - SNe and stellar winds are less effective at removing metals from the gas in deep potential wells \citep{Larson1974, Tremonti2004, Liang2016, Beverage2023}; ii) Massive quiescent galaxies are the oldest and most enriched in $\alpha$-elements, $\log([\mathrm{\alpha}/\mathrm{Fe}])$ \citep{Thomas2005, Spolaor2010, Conroy2014, McDermid2015, Beverage2023}. This is an indication of the relative enrichment due to core-collapse or delayed Type 1a SNe, directly probing the SFR timescales within the galaxy \citep{Matteucci1994, Rennehan2020}. 

Even from these two examples one can understand why the study of galactic chemical enrichment is a powerful probe that one can use in our numerical simulations to give us insight into the evolution of galaxies. Therefore, tracking the chemical enrichment and its stellar feedback counterpart as accurately as possible is a necessity.

In this paper, we incorporate a new stellar feedback and chemical enrichment model into the \texttt{SIMBA} cosmological simulation \citep{Dave2019}, which is named \texttt{SIMBA-C}. This is done to improve the simplified nature of the existing enrichment model, namely the instantaneous recycling of the metals model approximation. The instantaneous recycling model uses the assumption that all stars more massive than $1 \textup{M}_\odot$ die instantaneously, while all stars less massive live forever to be able to define the yield per stellar generation $y_{\mathrm{i}}$ and the returned fraction $R$ of the mass ejected into the interstellar medium (ISM) \citep{Matteucci2016}. We provide a more realistic cosmic chemical enrichment model in the simulation that improves the number of elements that can be formed and track these elements more accurately in the simulation. The simplified models tend to track only eleven elements (H, He, C, N, O, Ne, Mg, Si, S, Ca, and Fe), and neglect all other elements \citep{Dave2016, Dave2019}. Our new model also includes more types of stellar feedback events in addition to the event types already included. Finally, we also improve on the instantaneous recycling approximation that is common practise in simplified enrichment models, to be more self-consistent and to take into account the current state of the star particle within the simulation in its evolution. Using the method in \citet{Kobayashi2004}, we, therefore, treat the star particles within the simulation as evolving star particles rather than just allowing them to randomly experience stellar feedback processes to create the right amount of metals.

In this paper, Sec. \ref{sec: Simulation methodology} discusses the input physics of the \texttt{SIMBA-C} model, which is highly dependent on \texttt{SIMBA}, with significant modification in the chemical enrichment and feedback modules. In Sec. \ref{sec: Results}, we study the global galaxy properties, including stellar, star formation, and enrichment properties, particularly abundance ratios, which are substantially improved in \texttt{SIMBA-C} relative to \texttt{SIMBA}. Lastly, we describe our conclusions in Sec. \ref{sec: Conclusions}.

\section{Simulation methodology}\label{sec: Simulation methodology}

\subsection{\texttt{SIMBA}}\label{sec: Simba simulation}

Our new \texttt{SIMBA-C} simulations are based on the \texttt{SIMBA} simulation \citep{Dave2019}.  The \texttt{SIMBA} simulation is a large-volume cosmological simulation that uses the hydrodynamics + gravity solver \texttt{GIZMO} \citep{Springel2005, Hopkins2015, Hopkins2017}.  While we summarise the \texttt{SIMBA} model in this Section, we point the interested reader to \citet{Dave2019} for further details, while highlighting the changes made in \texttt{SIMBA-C}.  \texttt{GIZMO} evolves the hydrodynamic equations using the mesh-free finite mass (MFM) method and handles shocks using a Riemann solver without artificial viscosity.  It also preserves the mass within each fluid element at simulation time, thereby enabling simple tracking of gas flows.  Therefore, it marries the advantages of a particle-based code, such as adaptivity in space and time, with the precision of a Riemann solved-based mesh code \citep{Hopkins2015, Dave2019, Asensio2023}.

\texttt{SIMBA} handles radiative cooling and photoionization heating of the gas using the \texttt{GRACKLE-3.1} library \citep{Smith2016}, including metal cooling and non-equilibrium evolution of primordial elements.  The adiabatic and radiative terms are evolved together during a cooling sub-time-step which results in a more accurate and stable thermal evolution, particularly in the stiff regions of the cooling curve \citep{Smith2016, Dave2019}.  This model also includes self-shielding self-consistently during the simulation, based on the \citet{Rahmati2013}'s prescription in which the ionizing background strength is attenuated depending on gas density.  A spatially uniform ionizing background is assumed as specified by \citet{Haardt2012} but modified for self-shielding \citep{Dave2019}.

Star formation in \texttt{SIMBA-C} is modelled using a H$_{2}$-based star formation rate (SFR), where the H$_{2}$ fraction is computed based on the sub-grid model of \citet{Krumholz2011} based on the metallicity and local column density of H$_{2}$. The SFR is given by the H$_{2}$ density, and the dynamical time is scaled with the efficiency of star formation parameter $\epsilon_{*}$ \citep{Kennicutt1998}:
\begin{equation}
    \textup{SFR} =\frac{\epsilon_{*}\rho_{\mathrm{H_{2}}}}{t_{\mathrm{dyn}}}, \mbox{ }\epsilon_{*}=0.026.
\end{equation}
Compared to \texttt{SIMBA}, the only change in \texttt{SIMBA-C} is that we use the updated value for $\epsilon_{*}$ from \citet{Pokhrel2021}, rather than 0.02 as in \citet{Dave2019}.  This has no noticeable effect on our results.

\texttt{SIMBA}'s chemical enrichment model follows \citet{Oppenheimer2006} and tracks 11 elements (H, He, C, N, O, Ne, Mg, Si, S, Ca, and Fe) during the simulation, with enrichment tracked from Type II SNe, Type Ia SNe, and Asymptotic Giant Branch (AGB) stars. This uses \citet{Nomoto2006} for SNII yields, \citet{Iwamoto1999} for SNIa yields, and includes AGB star enrichment; see \citet{Oppenheimer2006} for more details. Type Ia SNe and AGB wind heating are included, along with interstellar medium (ISM) pressurization at a minimum level to resolve the Jeans mass in star-forming gas as described in \citet{Dave2016}.  This module is now completely different in \texttt{SIMBA-C}, as we describe in detail in Sec. \ref{sec: Chem5 Model}.

\texttt{SIMBA-C}, like \texttt{SIMBA}, models for star formation-driven galactic winds employ decoupled two-phase winds, with a mass loading factor as a function of stellar mass based on particle tracking results from the Feedback in Realistic Environments (FIRE) simulations by \citet{Angles2017b}. \texttt{SIMBA} uses the on-the-fly approximated friends-of-friends (FOF) finder applied to stars and dense gas as described in \citet{Dave2016} to compute galaxy properties such stellar mass, as well as circular velocities using a scaling based on the baryonic Tully-Fisher relation used to set the wind ejection speed.  While the mass loading scaling is unchanged, \texttt{SIMBA-C} differs from \texttt{SIMBA} in the velocity scaling, whose normalisation has been lowered from the original value of 1.6 to $0.85$; the latter value used in \texttt{SIMBA-C} matches the median value predicted by \citet{Muratov2015}.

In \texttt{SIMBA}, unlike in \texttt{SIMBA-C}, ejected winds are metal loaded, i.e. when a wind particle is launched, it extracts some metals from nearby particles to represent the local enrichment by the supernovae driving the wind.  The metallicity added to the wind particle is given by:
\begin{equation}
    \textup{d}Z = \frac{f_{\mathrm{SNII}}y_{\mathrm{SNII}}(Z)}{\mathrm{MAX}(\eta,1)},
    \label{eq. metal-loading winds}
\end{equation}
where $f_{\mathrm{SNII}}=0.18$ is the stellar mass fraction lost to supernovae (assumed to be instantaneous in \texttt{SIMBA}), $y_{\mathrm{SNII}}(Z)$ is the metal-dependent Type II SNe yields for each species, and $\eta$ is the wind mass loading factor.  The metal mass is subtracted from the nearby gas in a kernel-weighted manner. If there are not enough metals in the gas particles nearby, then $\mathrm{d}Z$ is appropriately reduced.

In \texttt{SIMBA}, Type Ia SNe and AGB stars also provide energetic feedback to the system but are delayed relative to the time of star formation \citep{Dave2016}. Delayed feedback is energetically subdominant relative to the processes driving the winds. \texttt{SIMBA} follows a concurrent (with Type II SNe) prompt component and a delayed component that emerges from stars after an age of $0.7\, \mathrm{ Gyr}$ for Type Ia SNe, modelled by \citet{Scannapieco2005}. Furthermore, \texttt{SIMBA} adds feedback energy from AGB stars to the surrounding gas, based on the model of \citet{Conroy2015}, in which AGB stellar winds are assumed to be thermalised with ambient gas \citep{Dave2016}. This module is now completely different in \texttt{SIMBA-C}, as we describe in detail in Sec. \ref{sec: Chem5 Model}.

\texttt{SIMBA-C} also includes black hole physics models, mostly following \texttt{SIMBA} with some updates.  In the original code, black holes are seeded into galaxies not already containing black holes whose stellar mass exceeds $M_* \gtrsim 5\times 10^{9}\, \mathrm{ M_{\mathrm{\odot}}}$. In \texttt{SIMBA-C}, we instead seed black holes when the galaxy is initially resolved ($M_* \gtrsim 6\times 10^{8}\, \mathrm{ M_{\mathrm{\odot}}}$), but we suppress black hole accretion exponentially with a factor $\exp(-M_{\rm BH}/10^6\, \mathrm{M_\odot})$, in order to roughly mimic the behaviour of star formation in dwarf galaxies suppressing black hole growth as described in \citet{Angles2017b, Hopkins2022}. This change alleviates the ``pile-up" of galaxies seen in \texttt{SIMBA}'s stellar mass function just above the $M_*$ where black holes are seeded, owing to a sudden onset of black hole feedback retarding growth; this will be shown for \texttt{SIMBA-C} in Sec. \ref{sec:global_properties}  At simulation time, both a dynamical mass and a physical black hole mass are tracked. The dynamical mass is inherited from the parent star particle (effectively the gas mass resolution). The physical mass is set to $M_{\mathrm{BH,seed}} = 10^4\, \mathrm{ M_{\mathrm{\odot}}} \, h^{-1}$ and allowed to grow via gas accretion.

For black hole growth, \texttt{SIMBA-C} continues to use \texttt{SIMBA}'s two-mode accretion model, namely the torque-limited accretion mode presented in \citet{Angles2017a} for the cold gas, while using Bondi-Hoyle-Lyttleton accretion for the hot gas as the second mode. The total large-scale accretion rate onto each black hole is given as the sum of the two modes, taking into account the conversion of matter into radiation:
\begin{equation}
    \dot{M}_\mathrm{BH} = \left(1-\eta\right) \times \left(\dot{M}_{\mathrm{Torque}} + \dot{M}_{\mathrm{Bondi}}\right),
\end{equation}
where $\eta=0.1$ is the radiative efficiency \citep{Yu2002}. $\dot{M}_\mathrm{Torque}$  follows the prescription discussed in \citet{Hopkins2011}, \citet{Angles2017a}, and \citet{Dave2019}, while the $\dot{M}_{\mathrm{Bondi}}$ follows the standard prescription discussed in \citet{Bondi1952}:
\begin{equation}
\begin{split}
    \dot{M}_{\mathrm{Torque}} &\equiv \epsilon_{T}f^{5/2}_{d}\bigg(\frac{M_{\mathrm{BH}}}{10^{8}\,\textup{M}_{\odot}}\bigg)^{1/6} \bigg(\frac{M_{\mathrm{enc}}(R_{0})}{10^{9}\,\textup{M}_{\odot}}\bigg)\\
    &\bigg(\frac{R_{0}}{100\,\textup{pc}}\bigg)^{-3/2}\bigg(1+\frac{f_{0}}{f_{\mathrm{gas}}}\bigg)^{-1}\textup{M}_{\odot}\,\textup{yr}^{-1}, \\
    \dot{M}_{\mathrm{Bondi}} &= \frac{4\pi G^{2}M^{2}_{\mathrm{BH}}\rho}{c_{\mathrm{s}}^{3}}.
\end{split}
\end{equation}
Like \texttt{SIMBA}, \texttt{SIMBA-C} limits the torque mode accretion rate to three times the Eddington accretion rate and the Bondi mode to the Eddington rate. 

The accretion energy is used to drive the black hole feedback that quenches galaxies. In \texttt{SIMBA-C} this is done by using a kinetic subgrid model for black hole feedback, along with X-ray energy feedback \citep{Dave2019}. The reason for this comes from the observed dichotomy in black hole growth modes that is reflected in their outflow characteristics (e.g. \citealt{Benson2009, Heckman2014}): A ``radiative mode'' at high Eddington ratios ($f_{Edd}\equiv \dot{M}_{BH}/\dot{M}_{Edd}\gtrsim$ few percent), in which AGNs are seen to drive multi-phase winds at velocities of $\sim 1000\, \mathrm{ km\, s}^{-1}$ that include molecular and warm ionised gas (e.g. \citealt{Sturm2011,Perna2017}); and a ``jet mode'' at low Eddington ratios, where AGNs, mostly drive hot gas in collimated jets that in clusters are seen to inflate super-virial temperature bubbles in surrounding gas (e.g. \citealt{McNamara2007, Jetha2008, Fabian2012}). According to \citet{Best2012, Dave2019}, this dichotomy also appears in radio jet activity (``high excitation'' vs. ``low excitation'' radio galaxies) above and below roughly a percent of the Eddington rate, with the former tending to be found in lower-mass, bluer host galaxies and the latter in massive early-types~\citep[see][]{Thomas2021}. 

\texttt{SIMBA-C}'s AGN feedback model is designed to directly mimic the energy injection into large-scale surrounding gas from these two modes, using purely bipolar feedback in the angular momentum direction of the black hole accretion radius (i.e. 256 nearest neighbours), with velocities and temperature taken as much as possible from AGN outflow observations.  In \texttt{SIMBA-C}, the velocity kick is given by 
\begin{equation}
    v_{w} = 500 + 500 \log_{10}\left(\frac{M_{\rm BH}}{10^6\,{\rm M_\odot}}\right)^{1/3} + v_{\rm jet},
\end{equation}
where $v_{\rm jet}$ is only included if the Eddington ratio $f_{\rm Edd}<0.2$, and is given by:
\begin{equation}
    v_{\rm jet} = 7000 \log_{10}\frac{0.2}{MAX(f_{\rm Edd},0.02)}\;{\rm km/s}.
\end{equation}

This differs from \texttt{SIMBA} only in the jet cap term, which was just $7000\, \mathrm{km/s}$, but now is roughly scaled with the halo escape velocity, so this cap scales up in the ability of jets to escape from halos in the group and cluster regime. This was originally proposed in the Three Hundred Cluster Zooms \citep{Cui2022}. However, it has a negligible effect on the mass scales covered in this work. 

Another change from \texttt{SIMBA} is the mass scale above which jets are allowed.  In \texttt{SIMBA}, this was $4\times 10^{7}\,\textup{M}_\odot - 6\times 10^{7}\,\textup{M}_\odot$, while in \texttt{SIMBA-C} it is $7\times 10^{7}\,\textup{M}_\odot - 1\times 10^{8}\,\textup{M}_\odot$. Moreover, in \texttt{SIMBA} this was implemented as a random value with a probability scaling from $0\to 1$ over that mass range, while in \texttt{SIMBA-C} each black hole particle is effectively assigned its own jet onset mass which it retains throughout the run.

To model the ejection as purely bipolar, the ejected gas elements are ejected in a positive and negative direction parallel to the angular momentum vector of the inner disk, typically the nearest $256$ gas particle neighbouring the black hole. \texttt{SIMBA} also included an energy input into the surrounding gas from X-rays of the accretion disk \citep{Dave2019} as motivated by \citet{Choi2012} assuming a radiative efficiency of $\eta = 0.1$.  \texttt{SIMBA} applies this heating only when the jet mode is active.

The above-described changes in \texttt{SIMBA-C} are a result of trial-and-error calibration, likely driven by the lower amount of metals produced by the \texttt{Chem5} model than by \texttt{SIMBA}'s instantaneous enrichment model. For example, this means that the metal cooling is less efficient, which motivates a later onset of jet feedback. The calibration is still done purely on stellar and black hole quantities, i.e. the stellar mass function evolution and the black hole mass-stellar mass relation, as was the case in \texttt{SIMBA}; no direct tuning was done based on the chemical enrichment model\footnote{The metallicity results were used as an indicator to calibrate the SFR. This was needed since the number of stars produced was lower than in \texttt{SIMBA} due to fewer amounts of metals predicted by the \texttt{Chem5} model, which led to a weaker metal cooling function, resulting in a stronger feedback system, creating fewer stars. Therefore, the feedback system was reconfigured (see Sec. \ref{sec: Chem5 Model}).}.

\subsection{The \texttt{Chem5} chemical enrichment model}\label{sec: Chem5 Model}
The primary new feature of \texttt{SIMBA-C} is the introduction of the new chemical enrichment model which we call \texttt{Chem5}. \texttt{Chem5} is the `version-5' of a self-consistent three-dimensional chemodynamical enrichment model developed by \citet{Kobayashi2004, Kobayashi2007, Taylor2014}\footnote{By self-consistent, we mean tracking the metal return following a detailed stellar evolution model with mass- and metal-dependent yields.}, with continued improvements by \citet{Kobayashi2011a} and \citet{ Kobayashi2020b, Kobayashi2020a}. Furthermore, the \texttt{Chem5} model has also been used in other studies such as \citet{Vincenzo2018a} and \citet{Ibrahim2023}. The model tracks all the elements from Hydrogen (H) to Germanium (Ge), thereby expanding on the existing list of elements within \texttt{SIMBA}. This model also introduces various physical processes associated with the formation and evolution of stellar systems. The physical processes that are included in the \texttt{Chem5} model include chemical enrichment and energy feedback from core-collapse supernovae (incl. hypernovae and `failed' supernovae), from Type Ia supernovae, and from stellar winds for stars of all masses including AGB and super-AGB stars.  We refer the reader to \citet{Kobayashi2020b, Kobayashi2020a} for a full description of each of these processes, including the yield table for each enrichment channel. Here, we will only be summarising the main concepts behind each process. 


%
%
%

\subsubsection{Stellar Winds}

As a star evolves, two components are taken into account: i) The metals that are processed through nucleosynthesis at the formation site to form new metals. ii) When a star is dying, stellar winds return a fraction or all of their unprocessed metals (locked-up in the star's envelope since its formation) to the ISM. According to \citet{Kobayashi2020a}, both components are included in the same nucleosynthesis table for AGB stars, while only processed metal yields are included in the supernovae yields (see Eq. 9 in \citealt{Kobayashi2000}). The amount of mass containing these metals that is ejected by the stellar winds in the \texttt{Chem5} model is given as:
\begin{equation}
    M_{\mathrm{wind}} = M_{\mathrm{init}}-M_{\mathrm{remnant}}-\sum _{i}p_{z_{i}m},
\end{equation}
where $M_{\mathrm{init}}$ is the initial mass of the star, $M_{\mathrm{remnant}}$ is the mass of the left-over remnant (e.g. black hole, neutron star, or white dwarf), while $p_{z_{i}m}$ is the nucleosynthesis yield of an element.

\subsubsection{Asymptotic Giant Branch Stars}
Depending on the metallicity, all stars with masses between $\sim 0.9-8\textup{ M}_{\odot}$ pass through the thermally-pulsing AGB phase \citep{Kobayashi2020a}. This phase, in particular, can lead to the enrichment of $^{12}$C on the surface of a star if the convective envelope is hot enough to sustain proton-capture nucleosynthesis. This happens because the He-burning shell becomes unstable during this phase, leading to processed metals from nuclear reactions in the core mixing with the envelope. Furthermore, this newly created $^{12}$C can then be converted into $^{14}$N by means of the CNO cycle (a \textit{primary} process). Similarly $^{23}$Na and Al can be created through the NeNa and MgAl cycles, respectively. Through stellar winds, these metals can be ejected from the star over its lifetime.

The ejected masses of each element are added to the yield tables. The new yields are calculated as the difference between the amount of the species in the winds and the initial amount in the envelope of the progenitor star. The initial abundances are set to the scaled proto-solar abundances calculated by \citet{Asplund2009}. Therefore, $p_{z_{i}m}$ could become negative, especially for H. It is therefore important to track the unprocessed metals to conserve mass.

\subsubsection{Super Asymptotic Giant Branch Stars}\label{sec: super AGBs}
The fate of high-mass stars, specifically stars with initial masses between $8-10\,\textup{M}_{\odot}$ (at $Z = 0.02$), is still uncertain \citep{Doherty2017, Kobayashi2020a}. It is accepted that the upper mass limit of the AGB stars, $M_{up,C}$, is defined as the minimum mass for the ignition of C. A hybrid C+O+Ne white dwarf (WD) can be formed when a star has a mass above this upper limit, due to neutrino cooling and contraction leading to ignition of C off-centre. This then propagates to the centre, but not completely all the way. However, below the upper mass limit, two different outcomes can occur. If the mass is $\lesssim 9\,\textup{M}_{\odot}$, the C ignition moves all the way to the centre, while it will undergo central C ignition if the mass is $\gtrsim 9\,\textup{M}_{\odot}$. Both of these outcomes will result in a degenerate O+Ne+Mg core, which could lead to an O+Ne+Mg WD, if the outer shell is lost through stellar winds. 

From this we can extract another upper mass limit, specifically for Ne: $M_\mathrm{up,Ne} \sim 9\,\textup{M}_{\odot}$. This in turn will lead to a new off-centre ignition, this time for Ne. This will then result in WDs of O+Ne+Mg or O+Ne+Fe. These processes and nucleosynthesis yields up to Ni have been included in the \texttt{Chem5} tables. They were collected from \citet{Doherty2013, Doherty2014}, and are scaled to the solar abundances from \citet{Grevesse1996}.

\subsubsection{Type Ia Supernovae}
The majority (more than 75\% \citealt{Kobayashi2020b}) of SNe Ia are believed to have occurred due to Chandrasekhar mass WDs. The \texttt{Chem5} model's nucleosynthesis yields for Type Ia are based on this fact. Therefore, because of the model's reliance on this, understanding of the evolution of accreting C+O WDs toward this mass is important. \citet{Kobayashi2020b} noted that two scenarios have been proposed: i) A double-degenerate (merging of two WDs). ii) A single degenerate, where the WD's mass grows via accreting mass from a binary companion, which is rich in H-matter. The \texttt{Chem5} model follows the latter scenario. The lifetime distribution function of the SNe Ia for this scenario in the \texttt{Chem5} model is calculated using:
\begin{equation}
\begin{split}
    R_{\mathrm{Ia}} = &b\int^{m_\mathrm{p,u}}_{\mathrm{max}[m_\mathrm{p,l},m_\mathrm{t}]}\frac{1}{m}\phi (m)dm\\ &
    \times \int^{m_\mathrm{d,u}}_{\mathrm{max}[m_\mathrm{d,l},m_\mathrm{t}]}\frac{1}{m}\psi (t-\tau_\mathrm{m})\phi_\mathrm{d}(m)dm, 
\end{split}
\end{equation}
where these integrals are calculated separately for the primary and secondary stars ($m_p$ and $m_d$, \citealt{Kobayashi2009}), taking into account the metallicity dependence of the WD winds \citep{Kobayashi1998} and the mass stripping effect on the binary companion stars \citep{Kobayashi2009}. 

A double peak is obtained, since the lifetime of the SNe Ia is determined by the companion stars. These peaks occur between: i) $\sim 0.1 -1\, \mathrm{Gyr}$ for main-sequence-WD systems, dominating in star-forming galaxies. ii) $\sim 1-20\, \mathrm{Gyr}$ for re-giant-WD systems -- dominating in early-type galaxies \citep{Kobayashi2009, Kobayashi2020b}. Furthermore, for the yields themselves, they are calculated with delayed detonations in Chandrasekhar-mass C+O WDs as a function of metallicity with solar-scaled initial compositions \citep{Kobayashi2020b}.

It must be noted that with recent observations of SNe Ia in galaxies and clusters, an empirical approach Delay Time Distribution (DTD -- see \citealt{Maoz2017}) rather than an analytical approach is often used to determine the SNe Ia rates in galaxy simulations. However, as shown in \citet{Kobayashi2020b, Kobayashi2023b}, this approach does not reproduce the elemental abundance observations in the Milky Way. The analytical formula can reproduce not only the elemental abundance observations but also the DTD. It is based on progenitor models, including the metallicity dependence. With this formula, it is also possible to include sub-Chandrasekhar-mass explosions of SNe Ia in the future.

\subsubsection{Core-collapse Supernovae}\label{sec: Core-collapse Supernovae}
According to \citet{Kobayashi2020a}, the \texttt{Chem5} model only uses nucleosynthesis yields from 1D calculations as presented in \citet{Kobayashi2006, Kobayashi2011b} for core-collapse SNe. The yields for massive star supernovae have been provided by three different groups (\citealt{Woosley1995, Nomoto1997a, Limongi2003}) and are constantly being updated. The reason for this is that the explosion mechanism is still uncertain for Type II, Ib, and Ic SNe. Furthermore, the Fe mass ejected by various different multidimensional simulations is not as large as observed, such as in \citet{Marek2009}. 

Some of the uncertainties in these yields are the reaction rates, mixing in stellar interiors, rotationally induced mixing processes, mass loss via stellar winds, and the formation of remnants (see \S 3.6 of \citealt{Kobayashi2020a}). Different solutions have been modelled to solve these uncertainties, such as the mixing fall-back model introduced by \citet{Umeda2002}, to mimic this multidimensional simulation phenomenon in 1D calculations. The \texttt{Chem5} model follows the approach where the quantities, such as the ejected explosion energy and $^{56}$Ni mass\footnote{$^{56}$Ni mass decays to $^{56}$Fe and forms most of the ejected Fe mass.} are determined independently through light curve and spectral fitting of an individual SN \citep{Kobayashi2006, Nomoto2013, Kobayashi2020a}, thereby constraining them with observations. This method was introduced in \citet{Kobayashi2006}.

As a result, hypernovae (HNe, \citealt{Kobayashi2020a}) are produced by the \texttt{Chem5} model. HNe are core-collapse SNe with masses between $M\geq 20\,\textup{M}_\odot$ and have explosion energies greater than $10\times$ that of a regular SN. They also produce more Fe and $\alpha$ elements. As for the nucleosynthesis yields for SNe II and HNe, they are provided separately as a function of the progenitor mass and metallicity. The fraction of HNe at a given time is uncertain and is set to a maximum of $\epsilon_{HN} = 0.5$ with declining values based on the metallicity dependency introduced in \citet{Kobayashi2011a}. It is important to note that the \texttt{Chem5} model does not include stellar rotation which is believed to be necessary to explain the observed N/O-O/H relation \citep{Frischknecht2016, Limongi2018}. According to \citet{Vincenzo2018b}, by using a more self-consistent cosmological simulation, the observed relation can be reproduced with inhomogeneous enrichment from AGB stars.

\subsubsection{`Failed' supernovae}
The upper limit of SNe II is not well known, because no progenitor stars are found at the locations of nearby SNe II with initial masses $M>30\,\textup{M}_\odot$. This has raised the question of whether massive SNe II can explode\footnote{Multidimensional simulations also seem to find it difficult to explode stars with $\gtrsim 25\,\textup{M}_\odot$ with the neutrino mechanism \citep{Janka2012}.}. To take this into account, \citet{Kobayashi2020a} introduced a new set of nucleosynthesis yields, called `failed' supernovae, at the massive end of SNe II while keeping the contributions from HNe. 

For the yields of `failed' supernovae, it is assumed that all CO cores fall onto black holes and are not ejected into the ISM in multi-dimensional timescales, since it is not long enough to follow this process. Therefore, the upper mass limit is treated as a free parameter while it stays the same for HNe. It is important to note that `failed' supernovae are not the same as `faint' supernovae, which are not included in the \texttt{Chem5} model.

\subsection{Implementation} \label{sec: Stellar feedback loop}
In this Section, we detail the integration of the \texttt{Chem5} chemical enrichment model and its various physical enrichment processes into \texttt{SIMBA}.

First, we send the information for every single star particle to the \texttt{Chem5} model at the beginning of every time step. The \texttt{Chem5} model treats star particles as an evolving stellar population that can eject thermal energy, mass, and heavy elements based on the initial and current properties of the stellar particle \citep{Kobayashi2004}. The contributions from all the stellar feedback channels within a given time step are determined. 

Two things must be noted: First, since a star particle represents an entire stellar population, it can undergo multiple events. If multiple events occur simultaneously, for instance, a SNe Ia and a AGB stellar feedback event, then the \texttt{Chem5} model adds the results for the two events together and returns a value for each of the three ejected variables, namely mass, metals, and energy. We distribute these ejecta to neighbouring gas particles in a kernel-weighted fashion. In \texttt{SIMBA}, it gets distributed to the nearest $64$ neighbouring gas particles. Specifically, for the injection of the metals, we use:
\begin{equation}
\begin{split}
    Z_{\mathrm{j,k}} = &\bigg(1 - \frac{\mathrm{d}M_{\mathrm{ej,in}}}{M_{\mathrm{j}}}\bigg)Z_{\mathrm{j,k}}+ \frac{\mathrm{d}M_{\mathrm{ej,in}}}{M_{\mathrm{j}}}Z_{\mathrm{ej,k}},
\end{split}
\end{equation}
where $Z_{\mathrm{j,k}}$ is the metallicity of the neighbouring gas particle, $\mathrm{d}M_{\mathrm{ej,in}}$ is the kernel-weighted ejected mass fraction, $M_{\mathrm{j}}$ is the gas particle's mass and $Z_{\mathrm{ej,k}}$ the ejecta metallicity for the star particle, while $k$ represents each element in the sample. After the distribution of these ejecta, $\mathrm{d}M_{\mathrm{ej,in}}$ is then removed from the star particle to ensure that the conservation of the metal mass/energy is followed. 

It is important to note that the energy ejection from SNe Type II in the \texttt{Chem5} model is not being distributed, since it is already being effectively employed by the star formation-driven wind model within \texttt{SIMBA}. The parameter $f_{\mathrm{SNII}}=0.18$ for \texttt{SIMBA}, is set to zero in \texttt{SIMBA-C} since this parameter represented an instantaneous mass loss due to Type II SNe within the first $\sim 30\, \mathrm{Myr}$, but now we are removing this instantaneous recycling approximation. 

Although the metal content influences metal cooling, we did not make any changes to the cooling function within \texttt{GRACKLE-3.1}, and therefore the cooling occurs exactly the same way as in \texttt{SIMBA}.

\subsection{Dust integration} 
We are not including the dust model from \texttt{SIMBA} as presented in \citet{Dave2019} in this update. The reason behind this is that the dust model influences the outcome of the individual elemental abundance, and for the purpose of this study, we want to test the \texttt{Chem5} model's impact on the \texttt{SIMBA} simulation. The dust model can run concurrently with the new model, but it has not been tested or calibrated to the new elemental abundances. We leave this to future work.

\subsection{Chabrier IMF} \label{sec: Chabrier IMF}
Since we are working with star particles that represent an entire stellar population, we need to adopt an Initial Mass Function (IMF) that describes the stellar mass distribution of this population. For consistency between \texttt{SIMBA} and the \texttt{Chem5} model, we use the Chabrier IMF (\citealt{Chabrier2003}), unlike \citet{Kobayashi2020a}. Specifically, we use a renormalised version of the Chabrier equation given by \citet{Romano2005}. The newly introduced Chabrier IMF is given as:

\begin{equation}
    \varphi_{\mathrm{Chabrier}} (m)=
    \begin{cases}
        A_{\mathrm{Chabrier}}\textup{e}^{(\log m -\log m_{\mathrm{c}})^{2}/2\sigma^{2}} &\text{if } m\leq 1\textup{ M}_\odot, \\
        B_{\mathrm{Chabrier}}m^{-1.3} &\text{if } m>1\textup{ M}_\odot;
    \end{cases}
\end{equation}

with $A_{\mathrm{Chabrier}} \simeq 0.79$ and $B_{\mathrm{Chabrier}} \simeq 0.22$ for our normalisation constants. Furthermore, we also use $m_{\mathrm{c}} = 0.079\,\textup{M}_\odot$ and $\sigma = 0.69$ as in \citet{Romano2005}. 

%
%
%


\subsection{Motivating the parameter changes from \texttt{SIMBA}}\label{sec: parameter changes}

\begin{table*}
	\centering
	\caption{Summary of all the differences between \texttt{SIMBA} (discussed in Sec. \ref{sec: Simba simulation}) and \texttt{SIMBA-C} (discussed in Sec. \ref{sec: Chem5 Model}), based on all the implementation and integration changes that were necessary (Sec. \ref{sec: Stellar feedback loop} to Sec. \ref{sec: Chabrier IMF}), and module/parameter changes due to calibrations (Sec. \ref{sec: parameter changes} and Sec. \ref{sec: metallicity distribution}).}\label{tab: Parameter changes}
	\begin{tabular}{lcc} 
		\hline
		\textbf{Changed ingredients} & \textbf{\texttt{SIMBA}} & \textbf{\texttt{SIMBA-C}}\\
		\hline
		Chemical enrichment model & Instantaneous recycling approximation & \texttt{Chem5} evolving chemical enrichment\\
        Stellar feedback types & Stellar winds, SNe Ia, AGB, SNe II & Stellar winds, SNe Ia, two AGBs, SNe II, HNe, `Failed SNe' \\
		Elements & H, He, C, N, O, Ne, Mg, Si, S, Ca, and Fe & H$\rightarrow $Ge\\
		Dust & Included & Excluded\\
        IMF & Chabrier & Chabrier $\sim$ changed from Kroupa\\
        Wind velocity scaling & $\alpha = 1.6$ & $\alpha = 0.85$ (now matching \citealt{Muratov2015})\\
        Black hole jet mass activation min & $4\times 10^{7}\,\textup{M}_\odot$ & $7\times 10^{7}\,\textup{M}_\odot$\\
        Black hole jet mass activation max & $6\times 10^{7}\,\textup{M}_\odot$ & $1\times 10^{8}\,\textup{M}_\odot $\\
        SNe II mass fraction & $f_{SNII} = 0.18$ & $ f_{SNII} = 0$\\
		\hline
	\end{tabular}
\end{table*}

Our initial attempt to incorporate the \texttt{Chem5} model into \texttt{SIMBA}, without recalibrating, significantly under-produced metals compared to the observational results in \citet{Yates2021} at all redshifts. We found that the metal cooling rates at low redshift are so inefficient that the stellar feedback model overheats all of the gas -- leading to fewer stars being produced, in turn creating fewer metals. Therefore, we found that it is essential to recalibrate \texttt{SIMBA}. 

First, we found that we must increase the minimum black hole jet mass threshold, which leads to fewer stellar feedback events occurring. Recall that \texttt{SIMBA} has a black hole mass range where the jet begins to operate, $M_\mathrm{min, jet}$, and slowly ramps up to maximum power at $M_\mathrm{max, jet}$. The mass range was kept within the same order of magnitude. Specifically, we changed this black hole jet activation parameter from $4\times 10^{7}\,\textup{M}_\odot - 6\times 10^{7}\,\textup{M}_\odot$ to $7\times 10^{7}\,\textup{M}_\odot - 1\times 10^{8}\,\textup{M}_\odot$. Our changes increased the SFRs of the simulated galaxies, which in turn increased the number of metals in the system.

However, our changes negatively affected the $L^*$ region of the $z=0$ galaxy stellar mass function (GSMF). We found that we needed to further recalibrate \texttt{SIMBA} by changing the wind velocity scaling. Recall that \texttt{SIMBA} stellar feedback relies on the results of \cite{Muratov2015} for wind mass loading $\eta$ and wind velocity $v_\mathrm{w}$. In \texttt{SIMBA}, the wind velocity followed:

\begin{equation}
\label{eq:simba_wind_velocity_scaling}
v_\mathrm{wind} = 1.6\,\bigg(\frac{v_\mathrm{circ}}{200 \, \mathrm{km} \, \mathrm{s}^{-1}}\bigg)^{0.12} + v_\mathrm{corr},
\end{equation}

\noindent where $v_\mathrm{circ}$ is the circular velocity of the galaxy, and $v_\mathrm{corr}$ is a correction term for the wind launch location compared to \cite{Muratov2015}.  The normalisation, $a = 1.6$, of this equation was set much higher in \texttt{SIMBA} compared to $a = 0.85$ in \cite{Muratov2015}.
By lowering this value from the default $a=1.6$ \texttt{SIMBA} value to the $a=0.85$ originally found in \cite{Muratov2015}, we improved the GSMF and cosmic star formation rate's history. Specifically, it increased $z=2$ SFR, while lowering it at $z=0$, resulting in a better match with the observations of \citet{Madau2014}. All of the above-mentioned changes are shown in Table \ref{tab: Parameter changes}.

\subsection{Runs and analysis}
In this paper, we have two simulations: i) The original \texttt{SIMBA} volume with the previous chemical enrichment module as a control, as well ii) Our \texttt{SIMBA-C} code including \texttt{Chem5}\footnote{The \texttt{MUSIC}-created initial conditions are the same for both simulations.}. Both runs consist of volumes of side-length $50 \mathrm{\, Mpc \, h^{-1}}$ with $512^{3}$ gas and $512^{3}$ dark matter particles down to $z = 0$.  Given that we are most concerned with enrichment in star-forming galaxies in this paper, the volume is sufficient to get a good sample of such objects.

Both runs begin at $z=249$ and follow a \citet{Planck2018} $\Lambda$CDM cosmology of $\Omega_{m}=0.3$, $\Omega_{\Lambda}=0.7$, $\Omega_{b} = 0.048$, and $H_{0} = 68 \mathrm{\, km \, s^{-1} \, Mpc^{-1}}$. We analyse the simulation outputs using a friends-of-friends galaxy finder to identify galaxies, assuming a spatial linking length of $0.0056$ times the mean inter-particle spacing. Galaxies and halos are cross-matched in post-processing using \texttt{Caesar}, a \texttt{yt}-based package\footnote{Caesar documentation can be found at \url{https://caesar.readthedocs.io/en/latest/}}. Galaxy finding is applied to all stars and black holes, as well as to all gas elements with a density above the minimum SF threshold density $n_{H} > 0.13\, \mathrm{H\, atoms\, cm}^{-3}$. Black holes are assigned to galaxies to which they are most gravitationally bound. The central black hole is considered the most massive black hole in the galaxy.

\section{Results}\label{sec: Results}

We first investigate the impact of the \texttt{Chem5} model on global galaxy properties by comparing \texttt{SIMBA-C} with \texttt{SIMBA}. The purpose of these comparisons is to identify whether the new model is able to reproduce the accuracy of galaxy populations that \texttt{SIMBA} obtained in \citet{Dave2019}, and secondly to determine whether \texttt{SIMBA-C} has improved on \texttt{SIMBA}'s minor discrepancies compared to observations.

\subsection{Stellar properties}\label{sec:global_properties}

\begin{figure*}
    \includegraphics[width=\textwidth]{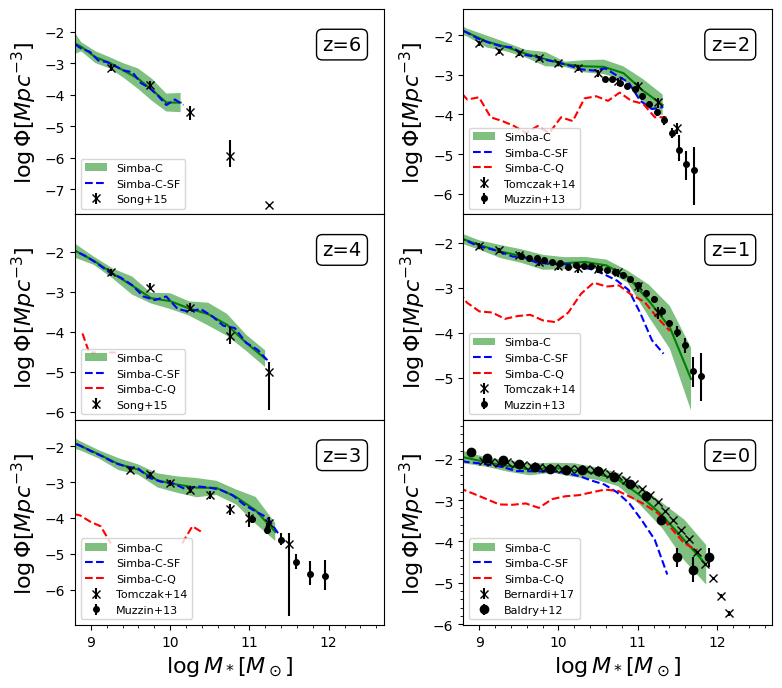}
    \caption{The evolution of the stellar mass function for the \texttt{SIMBA}-\texttt{C} simulation from $z=6 \rightarrow 0$, compared to different observations from \citet{Song2016} for $z= 6$ and $z=4$, \citet{Muzzin2013, Tomczak2014} for $z=3, 2, 1$, and \citet{Baldry2012,Bernardi2017} for $z=1$ and $z=0$. The red and blue dashed lines represent the median of the galaxy's sSFR and are divided into the two populations at $=10^{-1.8+0.3z}\,\mathrm{Gyr}^{-1}$, while the green band shows the combined results of all the star-forming and quenched galaxies within the \texttt{SIMBA}-\texttt{C} simulation.}
    \label{fig: mfssfr}
\end{figure*}

In this Section, we investigate the global scaling relationships and galaxy distributions through the GSMF, the black hole stellar mass relation, and the quenched fraction of the galaxy population.  We note that these properties have been tuned to reproduce observations in \texttt{SIMBA}, but we did not re-tune them for \texttt{SIMBA-C}.

The GSMF characterises the efficiency with which halos can convert baryons to stars \citep{Dave2011, Dave2019}. Since we are comparing the \texttt{SIMBA} model and the new \texttt{SIMBA-C} model, we use the same observational data as \citet{Dave2016} for comparison. 

In Fig. \ref{fig: mfssfr}, we show the evolution of the GSMF at $z=\{6,4,3,2,1,0\}$ for the \texttt{SIMBA}-\texttt{C} model. At $z=6$ and $z=4$, we used observational data based on the Cosmic Assembly Near-infrared Deep Extragalactic Legacy Survey (CANDELS) from \citet{Song2016}. For $z=\{3,2,1\}$, we used data from \citet{Tomczak2014} -- a combination of the CANDELS and the FourStar Galaxy Evolution Survey (zFOURGE) data -- as well as the results of \citet{Muzzin2013} from the Ultra Deep Survey with the Visible and Infrared Survey Telescope (UltraVISTA) within the Cosmic Evolution Survey (COSMOS). Lastly, for $z=0$, we use both the results of \citet{Bernardi2017} and \citet{Baldry2012} (part of the Galaxy and Mass Assembly -- GAMA project). 

Similarly to \citet{Dave2019}, we also show the subsets of the mass functions based on the SFR in Fig. \ref{fig: mfssfr}. These are shown as a blue line, representing the ``blue'' galaxies (star-forming galaxies), and a red line, representing the red and dead ``quenched'' galaxies. We use the specific SFR to distinguish between the two, namely the specific SFR (sSFR $= 10^{-1.8+0.3z}\, \mathrm{Gyr}^{-1}$). The combination of these two populations is our total stellar mass function, of which the median is plotted with the green line, and the green band represents the $1\sigma$ cosmic variance uncertainties on this calculation obtained by taking the variance over the GSMF in eight simulation octants.

From Fig. \ref{fig: mfssfr}, it is clear that on average \texttt{SIMBA}-\texttt{C} agrees quite well with the observed GSMF at all redshifts. 
For the higher redshift ranges, the blue star-forming galaxies follow the total mass function nearly identically, meaning that the contribution from quenched galaxies between $3\leq z \leq 6$ is almost non-existent.  This remains an issue with the \texttt{SIMBA} family of models, that they may not produce sufficient numbers of quenched massive galaxies at high redshifts~\citep{Sherman2020}. Furthermore, it is also clear that no massive star-forming galaxies are produced at these higher redshifts. This is related to the volume of the simulation box. At this volume, simulations produce very few massive galaxies, therefore it is improbable to form massive galaxies (star-forming or quenched) at earlier redshifts.

From $z=2$ to the present day, quenched galaxies start to contribute more, especially at higher masses ($M_{*} \gtrsim 10^{11}\,\mathrm{ M_{\odot}}$). This is quite evident for the blue star-forming galaxies at $z=1$ and $z=0$, where they drop steeply at $ M_{*}\sim10^{11.4}\,\mathrm{ M_{\odot}}$. The switch between high-redshift GSMF being dominated by star-forming galaxies, while quenched galaxies start to dominate at lower redshifts, is as expected, since galaxies start to quench due to AGN feedback from massive black holes combined with dropping gas accretion rates, and is broadly inferred from observations~\citep[e.g.][]{Faber2007}.

\begin{figure}
\includegraphics[width=\columnwidth]{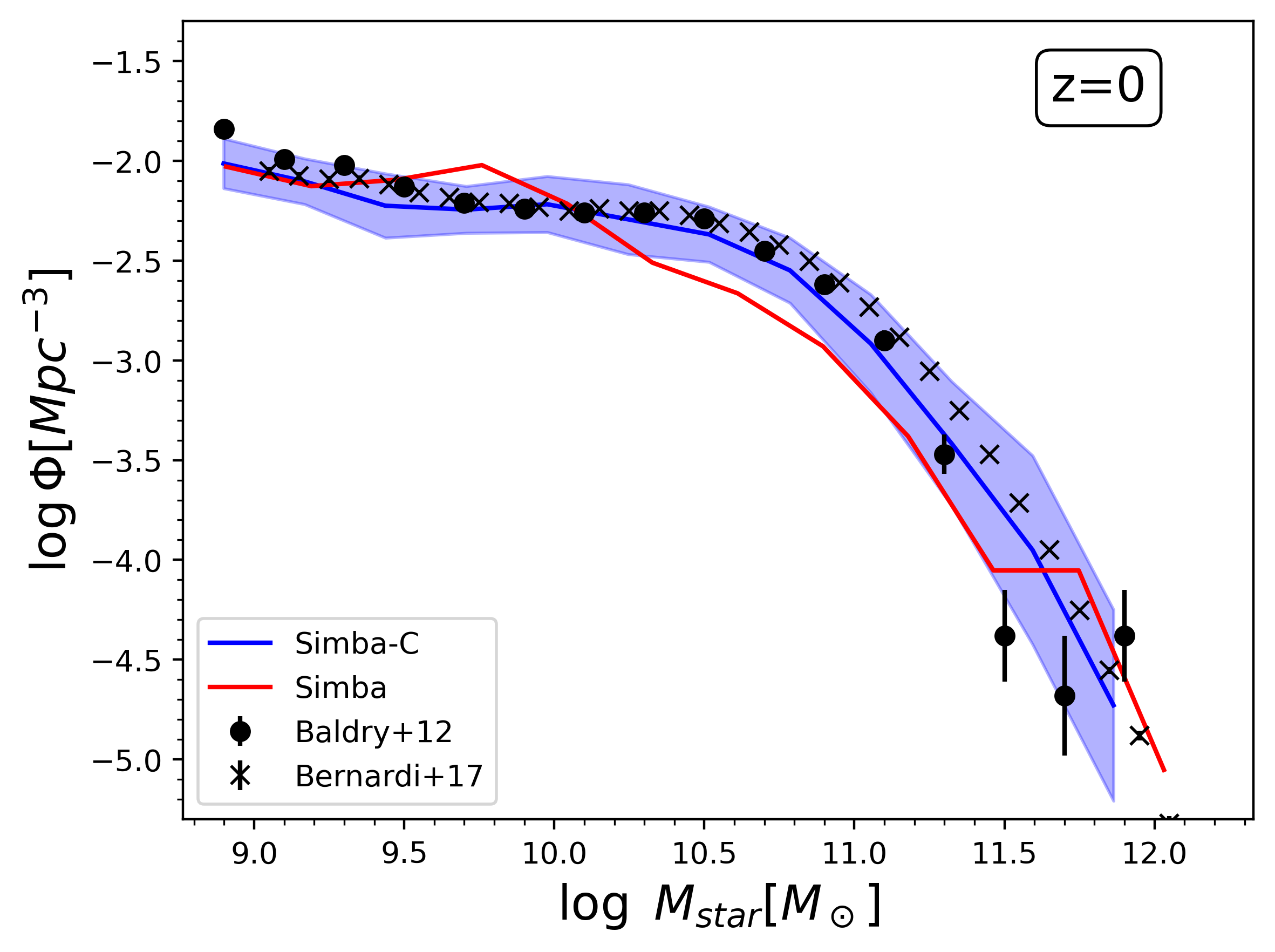}
    \caption{Comparison of the stellar mass function between the \texttt{SIMBA}-\texttt{C} simulation and the published version of \texttt{SIMBA} at $z=0$, compared to the same observations as in Fig. \ref{fig: mfssfr}. The \texttt{SIMBA}-\texttt{C} simulation median results (same results as the green band in the last panel of Fig. \ref{fig: mfssfr}) are shown by the blue line with its spread in the light blue band, while the red line displays the median \texttt{SIMBA} results for comparison.}
    \label{fig: msfunc}
\end{figure}

Fig. \ref{fig: msfunc} shows a comparison of the stellar mass function at $z=0$ between the \texttt{SIMBA} (red line) and \texttt{SIMBA}-\texttt{C} (blue band) results, compared to several recent observational determinations. Fig. \ref{fig: msfunc} clearly shows that the new model is in better agreement with the observational data.

Two areas stand out. The first is the evident improvement at $M_{*}\sim 10^{9.7}\,\mathrm{ M_{\odot}}$, where \texttt{SIMBA} shows a bump, while \texttt{SIMBA}-\texttt{C} shows no particular feature at that mass scale, which is more consistent with observations.  
Second, we see that in the mass range $10^{10.1}\,\mathrm{ M_{\odot}}\leq M_{*} \leq 10^{11.5}\,\mathrm{ M_{\odot}}$, on average, \texttt{SIMBA} predicts a GSMF value that is $\sim 0.3\,\mathrm{ dex}$ lower than the observations.  This was also seen in \citet{Dave2019}, where \texttt{SIMBA} undercuts the GSMF around the knee, an issue that is problematic for many galaxy formation models.  \texttt{SIMBA-C} shows a great improvement over \texttt{SIMBA} in this regard, with all observational results within the error, as well as the median of the simulation following the trend of the data almost identically, with a slight underestimate by $\sim 0.1\,\mathrm{ dex}$ for the lower mass end. This increase is the result of the change in the velocity scaling parameter (from $1.6$ to $0.85$ as described in Section \ref{sec: Stellar feedback loop}) necessary for the SFR due to the metal cooling function being inefficient. This change to the velocity scaling parameter is now more consistent with the result found in \citet{Muratov2015} using the FIRE simulations. Since \texttt{SIMBA} as a whole is calibrated and constrained to $z=0$, this is a valuable improvement, particularly for galaxies of the Milky Way size.

\begin{figure}
	\includegraphics[width=\columnwidth]{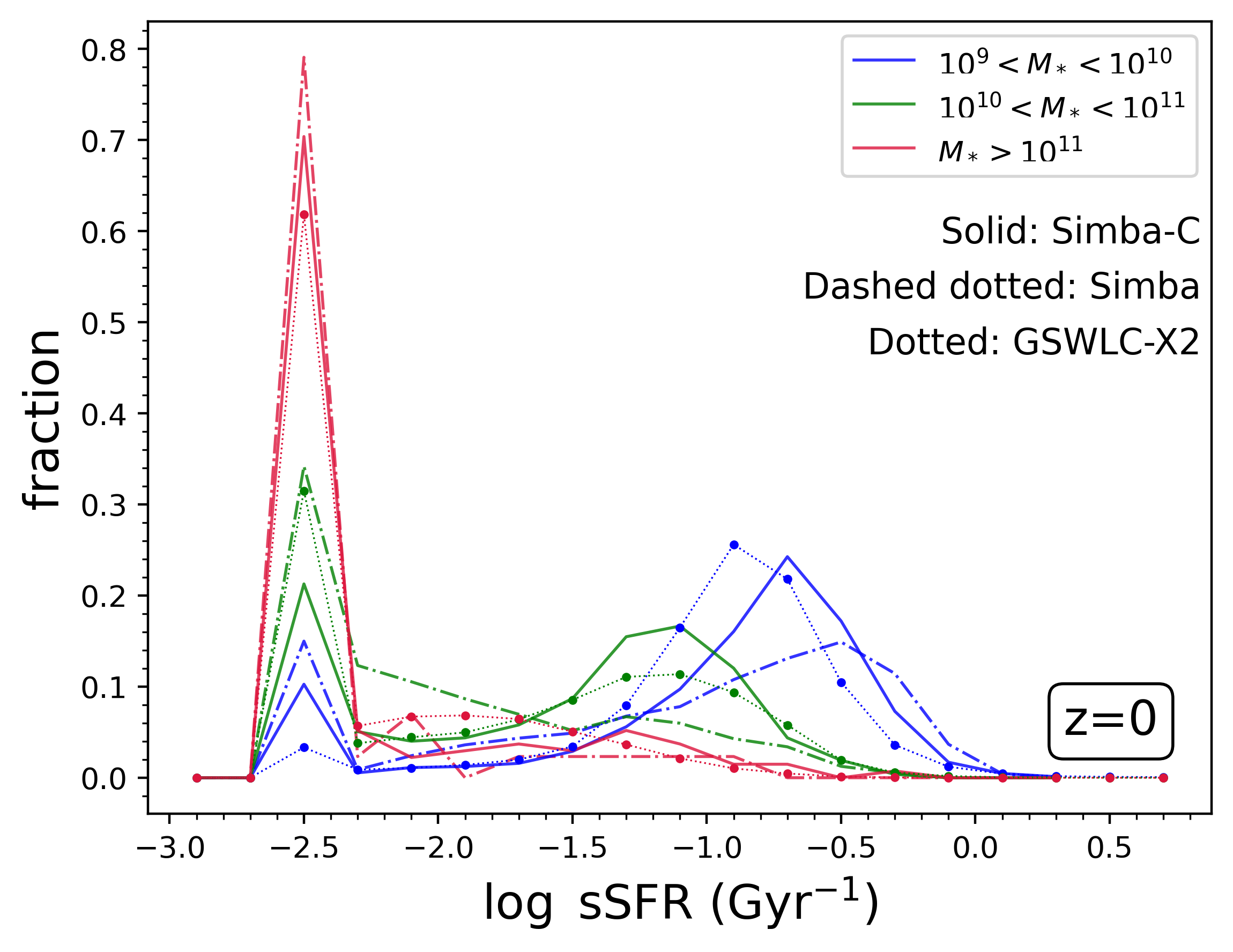}
    \caption{Histogram of sSFR in three bins of stellar mass. Solid lines show the results for \texttt{SIMBA}-\texttt{C}, and the dashed dotted lines show the \texttt{SIMBA} equivalent at $z=0$.  The observations are from the GSWLC-X2 catalog described in \citet{Salim2018}, with redshifts on average between $\sim 0.2>z\geq 0$ and shown with the dotted lines with a filled circle marker to distinguish them from their simulation counterparts. All galaxies with sSFR$<10^{-2.5}$ are placed in the lowest bin.}
    \label{fig: SFR_stellar_mass_relation_Hist_z0.png}
\end{figure}

Finally, we check that \texttt{SIMBA-C} is still quenching massive galaxies at $z=0$ in accordance with observations.  We show a histogram of the sSFR as a function of the mass of the galaxies at $z=0$ in Fig. \ref{fig: SFR_stellar_mass_relation_Hist_z0.png}. From the GALEX-SDSS-WISE LEGACY Catalog (GSWLC)-X2 \citep{Salim2018}, we use the master observations catalogue with the complete set of galaxies, which is different from the deep catalogue (which only uses a subset of the full catalogue) used in \citet{Dave2019}; this allows better statistics for massive galaxies, but the results are not significantly different. We binned the galaxies according to their stellar masses as follows: $10^{9}\,\mathrm{ M_{\odot}} < M_{*} \leq 10^{10}\,\mathrm{ M_{\odot}}$ (blue lines), $10^{10}\,\mathrm{ M_{\odot}} < M_{*} \leq 10^{11}\,\mathrm{ M_{\odot}}$ (green), and $ M_{*} > 10^{11}\,\mathrm{ M_{\odot}}$ (red). Similarly to \citet{Dave2019}, all sSFR$<10^{-2.5}$ are delegated to the lowest mass bin at that value. The observations are binned similarly and shown using dotted lines with a filled circle marker.

The overall bimodal distribution with a populous main sequence of blue galaxies, a similar number of quenched galaxies, and a dearth of galaxies in the green valley is present in both \texttt{SIMBA} runs, which is broadly consistent with the observations.  But there are significant differences in the details, between \texttt{SIMBA} and \texttt{SIMBA-C}.

Looking at each mass bin individually, starting with the lowest mass range, we see a noticeable improvement over \texttt{SIMBA} relative to the observations at the main-sequence peak.  \texttt{SIMBA-C} is much closer to the GSWLC-X2 data, although still with a slight offset from the higher sSFR.  This is noteworthy, considering the fact that \texttt{SIMBA} was already in good agreement with the low-mass star-forming galaxies. 

Another major area of improvement is in the green valley, particularly regarding intermediate mass galaxies (green lines), which dominate the green valley population.  In \texttt{SIMBA}, the green valley is not as empty as it should be, as also noted in \citet{Dave2019}, which results in less of a bimodal distribution than observed.  Conversely, \texttt{SIMBA-C} produces excellent agreement with observations around sSFR$\approx 10^{-1.5}-10^{-2}\,\mathrm{Gyr}^{-1}$, showing a clear drop relative to the populations at higher and lower sSFR within this mass bin. Hence, the impact of the new model appears to cause a more rapid transition of galaxies through the green valley within this key mass range, in better agreement with observations.

In the high mass range, \texttt{SIMBA}-\texttt{C} is also in better agreement with observations. For \texttt{SIMBA} the fraction of quenched high mass galaxies is $\sim 0.80$, while for \texttt{SIMBA}-\texttt{C} it is $\sim 0.70$, while for observations it is $\approx 0.65$. Although we see an improvement over \texttt{SIMBA}, \texttt{SIMBA-C} is still overproducing massive quenched galaxies relative to massive star-forming galaxies.

\begin{figure}
\includegraphics[width=\columnwidth]{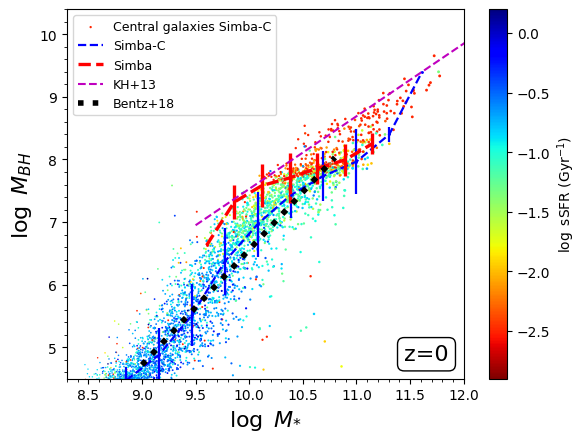}
    \caption{Comparison of the $M_{BH} - M_{*}$ relation at $z=0$ between the \texttt{SIMBA}-\texttt{C} simulation and \texttt{SIMBA}, and the observational results from \citet{Kormendy2013, Bentz2018}. We also include the central galaxy values from the \texttt{CAESAR} galaxy catalog for the \texttt{SIMBA}-\texttt{C} simulation, which were used to determine the median $M_{BH} - M_{*}$ value in its host galaxy and scaled their representing colour to their host galaxy's sSFR. The colour distribution corresponds to the blue being star forming galaxies, while red represents quenched galaxies.}
    \label{fig: mbhms}
\end{figure}

We also test the black hole mass to stellar mass ratio to see the impact of updating \texttt{SIMBA}'s black hole seeding and feedback modules. Fig. \ref{fig: mbhms} shows the comparison between the median \texttt{SIMBA} simulation results (red line) and the median \texttt{SIMBA}-\texttt{C} results (blue line) at $z=0$. We colour-coded the individual central galaxies contributing to the median based on their sSFR from the \texttt{SIMBA}-\texttt{C} model. We also use two sets of observational data, namely \citet{Kormendy2013} for comparison with the larger redder/quenched galaxies and \citet{Bentz2018} for the smaller blue/star-forming galaxies.

From Fig. \ref{fig: mbhms}, we see that \texttt{SIMBA}-\texttt{C} spans a much larger range in black hole masses and stellar masses of the host galaxies than \texttt{SIMBA}. Specifically, we have more blue star-forming galaxies following very closely the \citet{Bentz2018} observations, whereas \texttt{SIMBA} has a hard cut-off at $M_{*}\sim 10^{9.6}\,\mathrm{M_{\odot}}$. This owes to the earlier black hole seeding in \texttt{SIMBA-C}, and means that in the new version we are able to track black hole growth much earlier.  The agreement with the observed slope of \citet{Bentz2018} is adjusted by choosing the exponential accretion suppression factor.  However, \texttt{SIMBA-C} still does not reproduce the large scatter seen in observations at low $M_*$ in this relation (not shown).  Meanwhile, at the high $M_*$ end, we also find that our quenched galaxies follow the trend of \citet{Kormendy2013}, albeit somewhat lower; this is not significantly changed from \texttt{SIMBA}. The main take away from this is that the integration of \texttt{Chem5} model into \texttt{SIMBA} did not negatively change the black hole results, but also allowed tracking of black hole growth in lower mass galaxies.

\subsection{Metallicities and Abundance Ratios}\label{sec: metallicity distribution} 
The key change in \texttt{SIMBA-C} is the addition of the new chemical enrichment module.  To test this, we now investigate the metallicity content of our simulated galaxies. In this Section, we also discuss our recalibration of the stellar feedback strength through the wind launch velocity scaling. 

\begin{figure*}
    \includegraphics[width=\textwidth]{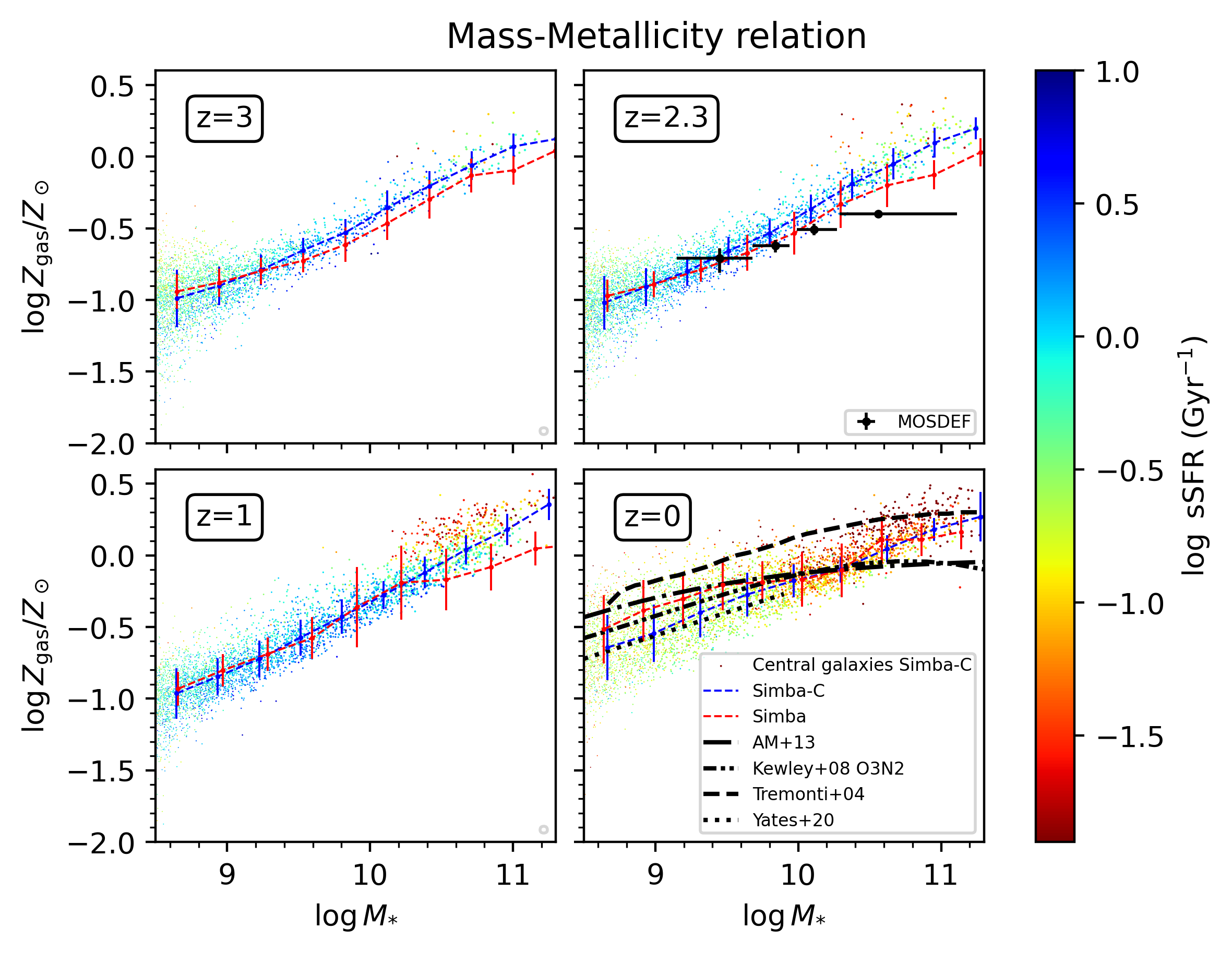}
    \caption{Comparison between the gas-phase mass-metallicity relation (MZR) from $z=3\rightarrow 0 $ between the median \texttt{SIMBA}-\texttt{C} results shown with the blue line based on its central galaxy values and median \texttt{SIMBA} results shown with the red line. We also compared both simulations with observations from \citet{Sanders2015} (MOSDEF) at $z=2$ and from \citet{Tremonti2004, Kewley2008, Andrews2013, Yates2020} at $z=0$. We also include the central galaxy values from the \texttt{CAESAR} galaxy catalogue for the \texttt{SIMBA}-\texttt{C} simulation and scaled their colour to their host galaxy's sSFR. The colour distribution corresponds to blue being star-forming galaxies, while red represents quenched galaxies.}
    \label{fig: mzr}
\end{figure*}

\subsubsection{Mass-metallicity relations}\label{sec: MZR}

First, we investigate the stellar mass -- gas-phase metallicity relation (MZR) and the stellar mass -- stellar metallicity relation. In Fig. \ref{fig: mzr}, we show the evolution of the MZR from $z=3$ to $z=0$. We colour-coded the central galaxies for the \texttt{SIMBA}-\texttt{C} model according to their sSFR and plot the median with a blue line, while showing the median \texttt{SIMBA} result with a red line. We use the observations from \citet{Sanders2015} as part of the Multi-Object Spectrometer for Infra-Red Exploration Deep Evolution Field (MOSDEF) survey at $z=2.3$. For $z=0$, we used the results of \citet{Tremonti2004, Kewley2008} and \citet{Andrews2013}, all of whom used the Sloan Digital Sky Survey (SDSS) observations, and \citet{Yates2020} using the Mapping Nearby Galaxies at APO (MaNGA) sample\footnote{The results from \citet{Kewley2008} are the same results as those of \citet{Tremonti2004}, but have been refitted using `strong line metallicities, where we use the calibration of $O_{3}N_{2}$.}. The total gas-phase metallicity was calculated as an SFR-weighted average of all gas particles in a galaxy, normalised to the solar value of 0.0134 \citep{Asplund2009}. The reason for using an SFR-weighted average is to compare to observations. In observations, (gas) metallicities are measured from nebular emission lines that arise in star-forming regions. Therefore, we only observe the metallicity of the gas that is star-forming. A simple way to account for this is to weigh the metallicity by the SFR of each particle.

Fig. \ref{fig: mzr} shows that the MZR of the two simulations are very similar between $z=3$ and $z=1$, although there are some differences emerging at higher masses between \texttt{SIMBA} and \texttt{SIMBA}-\texttt{C}, albeit not significant. The new model gives a higher total metallicity for the massive galaxies. The most interesting difference between the two simulations is at $z=0$. The median of the new model is on average lower than that of the \texttt{SIMBA} instantaneous model by $\sim 0.1-0.2\,\mathrm{ dex}$. This difference is mostly due to the calibration process explained in Sec. \ref{sec: Chem5 Model}. The initial tests showed a very low metallicity count, leading to fewer stars forming. By lowering the strength of the feedback system we were able to produce more stars and higher amounts of metals. However, this came at the cost of the GSMF, which we solved with the wind velocity scaling. Therefore, through trial and error, the best combination between these different components was determined, leading to slightly lower MZR values.

\begin{figure*}
    \centering
    \begin{minipage}{.5\textwidth}
        \centering
        \includegraphics[width=\linewidth]{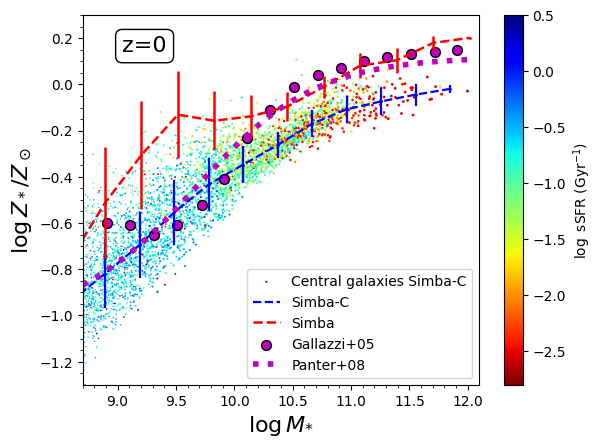}
    \end{minipage}%
    \begin{minipage}{.5\textwidth}
        \centering
        \includegraphics[width=\linewidth]{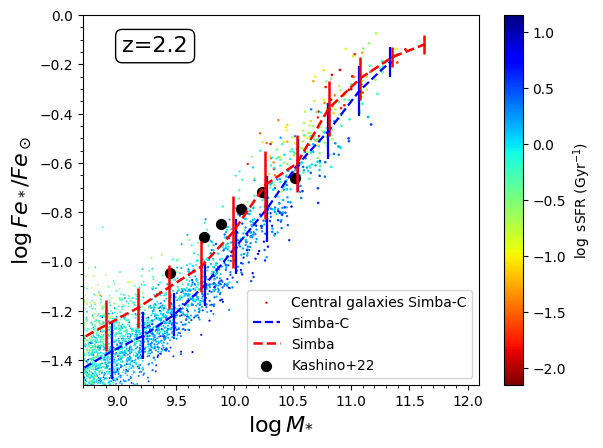}
    \end{minipage}
    \caption{Comparing the stellar mass-metallicity relation at $z=0$ using the total metallicity abundance (left panel), and at $z= 2.2$ using the iron abundance (right panel) between the median \texttt{SIMBA}-\texttt{C} results shown with the blue line based on its central galaxy values and median \texttt{SIMBA} results shown with the red line. We also compared both simulations with observations from \citet{Gallazzi2005, Panter2008} at $z\approx 0$ and \citet{Kashino2022} at $z\approx 2.2$. We also include the central galaxy values from the \texttt{CAESAR} galaxy catalogue for the \texttt{SIMBA}-\texttt{C} simulation and scaled their colour to their host galaxy's sSFR. The colour distribution corresponds to blue indicating star-forming galaxies, while red represents quenched galaxies.}
    \label{fig: stellar mass metallicity}
\end{figure*}

Next, we investigate the stellar mass-metallicity relation, shown in Fig. \ref{fig: stellar mass metallicity}. We plot the relation at $z=0$ and $z=2.2$ to directly compare with observations from \citet{Gallazzi2005} who used SDSS-DR2 data, \citet{Panter2008} who analysed the original SDSS data, and \citet{Kashino2022} who used the far-ultraviolet spectra from the high redshift zCOSMOS-deep survey. We again show the comparison plot between the median \texttt{SIMBA} results with a red line, and the median \texttt{SIMBA}-\texttt{C} results shown with a blue line. We also show the sSFR colour-coded distribution of central galaxies, as in Fig. \ref{fig: mzr}. The total and iron stellar metallicity was calculated as a SFR-weighted average of all star particles in a galaxy, normalised to the solar value of 0.0134 (\textit{Z}) and $1.31\times 10^{-3}$ (\textit{Fe})  \citep{Asplund2009}.

The stellar mass -- metallicity results are mostly consistent with the MZR results. The \texttt{SIMBA}-\texttt{C} results are overall $\sim 0.1-0.2\, \mathrm{ dex}$ lower than the \texttt{SIMBA} results. For $z=0$ in the lower mass range, the new model matches the observational results, especially \citet{Panter2008}, more accurately than \texttt{SIMBA}, while at higher masses \texttt{SIMBA} fares better in this regard. On average, \texttt{SIMBA}-\texttt{C} is more successful in following the general trends of the results of \citet{Panter2008}. As for the \citet{Gallazzi2005} results, neither \texttt{SIMBA} or \texttt{SIMBA}-\texttt{C} accurately reproduced the ``s''-shaped trend. 

On the other hand, at $z=2.2$, both the \texttt{SIMBA} and \texttt{SIMBA-C} simulation are overall more consistent with the observations than the $z=0$ result, especially \texttt{SIMBA}. This further confirms that both simulations match the observations, but that the overall trends obtained for both sets of observations are not fully reproduced in the simulations at $z=0$ and that further investigation of stellar evolution is necessary.

This inconsistency with the trends leads to a discrepancy between the gas-phase MZR and the stellar mass -- metallicity, in particular the steepness of the slope at $z=0$. As mentioned, for the gas-phase mass-metallicity (Fig. \ref{fig: mzr}), we obtain a steeper slope for both simulations compared to observations for higher stellar-mass galaxies, while this is not true for the stellar mass-metallicity (Fig. \ref{fig: stellar mass metallicity}). Here both simulations obtained a more flat slope compared to the observations, although \texttt{SIMBA} is more comparable to the observations in this mass range. However, this is a discrepancy between the based gas phase (having a steeper slope\footnote{This steeper slope is also seen for the \texttt{SIMBA-C} simulation in Fig. \ref{fig: O_H mzr}.}) and the stellar particles in the simulations (having a more flat slope) might not be related to the same problem.

Possible explanations include: i) For the gas-phase MZR, the steepness is set by the relationship between a mass loading factor and $M_{*}$ (see \citealt{Finlator2008, Dave2012}). Therefore, the assumed mass loading factor scaling with $M_{*}$ (taken from the \texttt{FIRE} simulation) might be too strong. ii) also for the gas-phase MZR, at higher masses the slope is sensitive to the type of massive galaxies produced in the simulation. Higher mass star forming galaxies have lower total stellar metallicities than the quenched massive galaxies. Therefore, more massive star-forming galaxies would flatten the slope. This corresponds to our GSFM (Fig. \ref{fig: mfssfr}) where we did not obtain any star-forming massive galaxies, which resulted in our MZR slope being dominated by the higher total stellar metallicity-valued quenched massive galaxies leading to an increased in the MZR slope. iii) For stellar mass -- metallicity, the inaccuracy might be due to using a mass-weighted metallicity for the star particles, whereas in observations it is measured from stellar absorption lines (see \citealt{Zahid2013} where they study empirical stellar metallicity models), which can be dominated by AGB stars in quenched galaxies. This might introduce some bias. More research is required to solve this discrepancy.

\subsubsection{Chemical abundances}\label{sec: chemical abundances}
In this Section, we present the key elemental abundance ratios used to study galactic chemical evolution, to investigate if \texttt{SIMBA-C} is an improvement over \texttt{SIMBA}'s previous instantaneous recycling of the metals model assumption. 

\begin{figure}
    \includegraphics[width=\columnwidth]{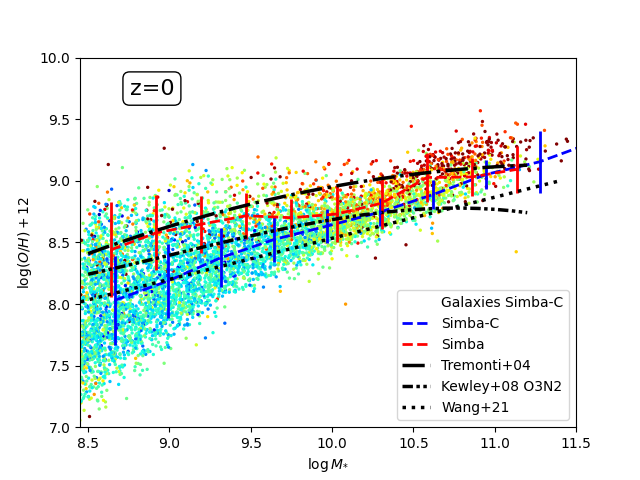}
    \caption{Comparing the gas-phase mass-oxygen abundance relation between the median \texttt{SIMBA}-\texttt{C} results shown with the blue line based on its central galaxy values and median \texttt{SIMBA} results shown with the red line. We also compared both simulations to observations from \citet{Tremonti2004, Kewley2008} using the SDSS catalog and \citet{Wang2021} using the spectroscopic surveys of MAD and MaNGA at $z=0$. We also include all of the galaxy values from the \texttt{CAESAR} galaxy catalog for the \texttt{SIMBA}-\texttt{C} simulation and scaled their colour to their host galaxy's sSFR. The colour distribution corresponds to blue indicating star forming galaxies, while red represents quenched galaxies. We use same the solar metallicity normalization of $8.69$ for $\log \mathrm{O}/\mathrm{H} +12$ used in \citet{Tremonti2004, Asplund2009, Grevesse2010}.}
    \label{fig: O_H mzr}
\end{figure}

In Fig. \ref{fig: O_H mzr}, we show the gas-phase oxygen abundance -- stellar mass ratio at $z=0$ (as a proxy for the MZR). We also include the observational results from \citet{Tremonti2004, Kewley2008}, who used the SDSS catalog data release, and the results from \citet{Wang2021}'s spectroscopic data from MUSE Atlas of Disks (MAD) and MaNGA.

\texttt{SIMBA-C} produces a lower normalization of the gas-phase oxygen abundance relation than \texttt{SIMBA}. This is similar to the lower MZR for both the gas-phase and the star particles in Fig. \ref{fig: mzr} and \ref{fig: stellar mass metallicity}. It is interesting to note that at lower galaxy masses, the two simulations follow the two separate observational results. At higher masses, the results converge. In this comparison, the slope of \texttt{SIMBA} is flatter than that of \texttt{SIMBA-C}, unlike in the previous section where both behaved similarly. This can be part of the slope discrepancy discussion in Sec. \ref{sec: MZR}, or part of a well-known problem, where the gas-phase mass-oxygen abundance relation MZR is known to differ by as much as $\sim 0.7\, \mathrm{ dex}$ and is not well constrained. A full discussion by \citet{Kewley2008}, using the fourth release of the SDSS data catalog, shows that \citet{Tremonti2004} seems to be on the higher end of the mass-oxygen abundance relation. This, together with the fact that \citet{Wang2021} uses newer observational results and obtains a lower mass-oxygen abundance relation, indicates a lower value would be more accepted as predicted by \texttt{SIMBA}-\texttt{C} even though it obtained a steeper slope.

\begin{figure*}
    \centering
    \begin{minipage}{.5\textwidth}
        \centering
        \includegraphics[width=\linewidth]{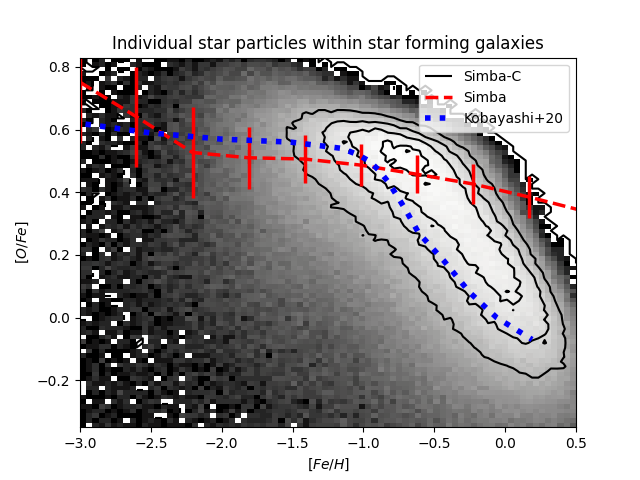}
    \end{minipage}%
    \begin{minipage}{.5\textwidth}
        \centering
        \includegraphics[width=\linewidth]{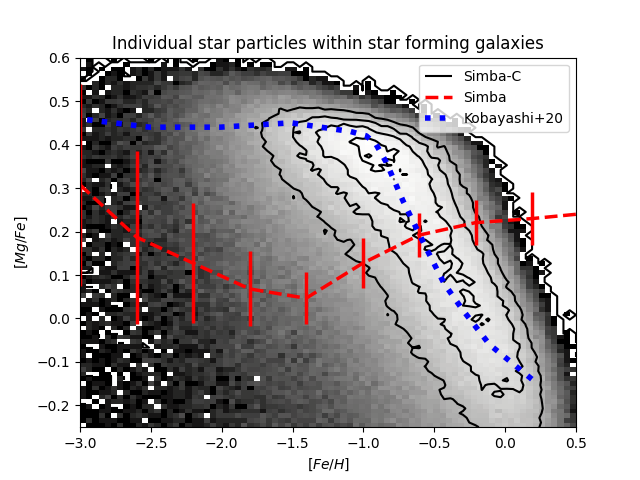}
    \end{minipage}
    \caption{The chemical abundance ratio plots for $[\mathrm{O}/\mathrm{Fe}]$ vs $[\mathrm{Fe}/\mathrm{H}]$ (left panel) and $[\mathrm{Mg}/\mathrm{Fe}]$ vs $[\mathrm{Fe}/\mathrm{H}]$ (right panel) for stars within Milky-Way like galaxies. The abundance ratios for the \texttt{SIMBA-C} simulation are shown with contour lines to show the majority of the values. We also included the theoretical model values from \citet{Kobayashi2020a} (blue dotted line), as well as the median \texttt{SIMBA} results (red line).}
    \label{fig: elemental abundances}
\end{figure*}

In the left panel of Fig.~\ref{fig: elemental abundances}, we show the oxygen abundance ratio trend $[\mathrm{O}/\mathrm{Fe}]$ as a function of metallicity $[\mathrm{Fe}/\mathrm{H}]$ and, in the right panel, we show the magnesium abundance trend $[\mathrm{Mg}/\mathrm{Fe}]$\footnote{We are using the notation $[\mathrm{X}/\mathrm{Y}]$ to represent the ratio between two elements that have been normalised to solar abundances, namely $\log ((X/Y)/(X/Y)_{\mathrm{solar}})$.}. The 2D histograms in the background of both panels show the respective abundance ratios of all the star particles in the \texttt{SIMBA-C} simulation that are in star-forming galaxies of masses $5.0\times 10^{10}\,\mathrm{ M_{\odot}} < M_{*} \leq 7.0\times 10^{10}\,\mathrm{ M_{\odot}}$. Additionally, the long-dashed line shows the median result of \texttt{SIMBA} (with the same selection), while the short-dashed line shows the reproduced results from \cite{Kobayashi2020a}.  By selecting only these star particles, we try to restrict our results to mimic the results from \citet{Kobayashi2020a}, where they use the systematic nonlocal thermodynamic equilibrium (NLTE) abundances from F and G dwarf stars in the solar neighbourhood (stars within the Milky Way) from observational results of \citet{Zhao2016}, to calibrate the original \texttt{Chem5} model. However, this is not a direct comparison since stars from the solar neighbourhood are not representative of all stars within Milky-Way like galaxies, but it does give us insight into how each simulation behaves relative to a model calibrated specifically to observations.

The left panel of Fig. \ref{fig: elemental abundances} shows that the observational data have two segments: i) a plateau with a slight negative slope at $[\mathrm{O}/\mathrm{Fe}] \sim 0.6$ for $[\mathrm{Fe}/\mathrm{H}]$ $\lesssim-1$, and ii) a sharp decline from $[\mathrm{O}/\mathrm{Fe}]$ $\sim 0.6$ to $[\mathrm{O}/\mathrm{Fe}]$ $\sim 0.0$ between $-1 \lesssim [\mathrm{Fe}/\mathrm{H}] \lesssim 0.3$. This well-known pattern (see \citealt{Wallerstein1962}, for example) can also be seen in the magnesium abundance ratio plot on the right-hand side panel, as predicted by \citet{Zhao2016}, who stated that all $[\mathrm{\alpha}/\mathrm{Fe}]$ ratios will show this pattern. The reason for this pattern is that in the early stages of galaxy formation, only SNe II/HNe contribute, leading to the $[\mathrm{O}/\mathrm{Fe}]$ plateau across a wide range of $[\mathrm{Fe}/\mathrm{H}]$ values \citep{Kobayashi2020a}. However, at $[\mathrm{Fe}/\mathrm{H}]$$\sim -1$, SNe Ia begins to occur and produces more iron relative to the $\alpha$ elements. The delayed enrichment of SNe Ia then results in the decreasing $[\mathrm{\alpha}/\mathrm{Fe}]$ ratio. \citet{Kobayashi2020a} incorporated these different interactions, based on the type of SNe, into the \texttt{Chem5} model to ensure that this characteristic pattern will emerge.

In Fig. \ref{fig: elemental abundances}, we also show the theoretical results of the \texttt{ Chem5} model (blue line) for the abundance of oxygen and magnesium calculated and calibrated in \citet{Kobayashi2020a}. By comparing to the theoretical results, we can validate the \texttt{SIMBA}-\texttt{C} model results to the original published \texttt{Chem5} model that is specifically tested against these abundance ratio trends.

From Fig. \ref{fig: elemental abundances}, we can see that the \texttt{SIMBA}-\texttt{C} simulation (the contour lines) accurately predicts the abundance ratio for both oxygen and magnesium. We have a small negative slope plateau and a clear decrease for both between $-1 < [\mathrm{Fe}/\mathrm{H}] < 0$. The decrease seen in the magnesium abundance follows the observations more closely than the oxygen abundance. As for the plateau, both elements have slopes similar to the observations and the theoretical \texttt{Chem5} model predictions, but unfortunately, the majority of our results do not match the very low $[\mathrm{Fe}/\mathrm{H}]$ values. However, the plateau still resembles this trend (as expected) but is not part of the majority of the results. This might be due to using all of the stars in Milky-Way-like galaxies and not only solar neighbourhood stars as in the original \texttt{Chem5} model's calibration. 

The \texttt{SIMBA} results are shown in Fig.~\ref{fig: elemental abundances} as a long-dashed line representing the median of the distribution at a given $[\mathrm{Fe}/\mathrm{H}]$.  It is clear that \texttt{SIMBA} cannot reproduce the trend of the abundance ratio, most likely due to its use of the instantaneous recycling approximation.  Therefore, the \texttt{SIMBA-C} simulation is a major improvement over the original \texttt{SIMBA} simulation in that we are now able to accurately predict the abundance ratio trends in Milky Way-like galaxies.

\begin{figure}
    \includegraphics[width=\columnwidth]{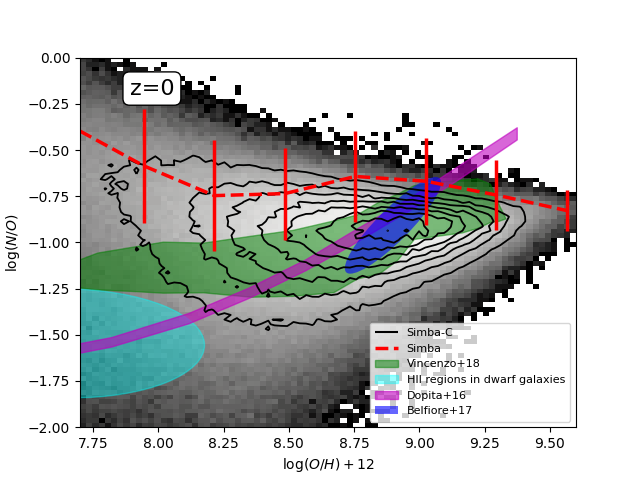}
    \caption{The chemical abundance ratio plot for $\log \mathrm{N}/\mathrm{O}$ vs $\log \mathrm{O}/\mathrm{H} +12$ for gas regions within star-forming galaxies with stellar masses between $10^{9} - 10^{11.5}\, \textup{M}_\odot$. The abundance ratios for the \texttt{SIMBA}-\texttt{C} simulation are shown with contour lines, while the median \texttt{SIMBA} values are shown using the red line. For observations, we include the observations presented in \citet{Vincenzo2018a} (green), from \citet{Belfiore2017} using the MaNGA spectroscopic results (blue), and HII regions in blue diffused dwarf galaxies consisting of various different observations from \citet{Berg2012, Izotov2012, James2019} (cyan), and the calculated observational average from \citet{Dopita2016} (magenta). The \citet{Dopita2016} observational average uses observations from both stellar and nebula sources \citep{Izotov1999, Israelian2004, Spite2005, Nieva2012}.}
    \label{fig: NO_vs_OH}
\end{figure}

Finally, in Fig. \ref{fig: NO_vs_OH} we show the $\mathrm{N}/\mathrm{O}$ abundance ratio as a function of the $\log \mathrm{O}/\mathrm{H} +12$ metallicity for individual gas particles within star-forming galaxies with stellar masses between\footnote{We chose this stellar mass range based on the \citet{Belfiore2017} MaNGA observations to allow for direct comparison. It also includes the galaxy stellar mass range used in \citet{Vincenzo2018a}.} $10^{9}-10^{11.5}\,\textup{M}_\odot$. We use the three observational data sets presented in \citet{Vincenzo2018a} as well as their own results (green). These include HII regions in diffused blue dwarf galaxies from \citet{Berg2012, Izotov2012, James2019} (cyan), a higher mass galaxies gas-phase study using the MaNGa spectroscopic survey from \citet{Belfiore2017} (blue) as well as the calculated average from various different stellar and nebulae sources from \citet{Izotov1999, Israelian2004, Spite2005, Nieva2012} as presented in \citet{Dopita2016} (magenta)\footnote{It should be noted that \citet{Dopita2016} also includes individual stellar sources in their average and should therefore be compared with gas-phase metallicity as a function time, unlike integrated stellar abundances of galaxies. Furthermore, they also use the solar normalised value of $\log \mathrm{O}/\mathrm{H} +12$ as $9.05$, which is different from the solar normalisation of $8.69$ from \citet{Tremonti2004, Asplund2009, Grevesse2010}.}.

Fig. \ref{fig: NO_vs_OH}, shows a clear difference between \texttt{SIMBA} and the \texttt{SIMBA}-\texttt{C} simulation, with the latter obtaining a lower $\log \mathrm{N}/\mathrm{O}$ value for the region of highest values concentration as a function of $\log \mathrm{O}/\mathrm{H} +12$. The \texttt{SIMBA}-\texttt{C} simulation results are in better agreement with observations. With lower values of $\log \mathrm{N}/\mathrm{O}$, we found that the highest concentration of values within the \texttt{SIMBA-C} coincides with the MaNGA result. This is significant as we limit our results to the stellar mass range of the MaNGA survey. Furthermore, this leads rather elegantly to the fact that the MaNGA survey excludes dwarf galaxy stellar mass ranges (up to a few $10^{9}\,\textup{M}_\odot$ \citealt{Revaz2018}). Since the \texttt{SIMBA-C} result is based on the MaNGA survey's stellar mass range, it is interesting to see that the concentrated levels decrease towards the \citet{Dopita2016} calculated observational average result, but does not quite reach the dwarf galaxy's $\log \mathrm{N}/\mathrm{O}$ vs $\log \mathrm{O}/\mathrm{H} +12$ values\footnote{We note that when we include dwarf galaxies, our concentration levels did in-fact include the dwarf galaxy region, but shifted our highest concentration of values away from the MaNGA results, while also increasing the spread/uncertainty of our results.}.

From these comparisons, we can deduce that \texttt{SIMBA-C} shows various improvements over \texttt{SIMBA}, with \texttt{SIMBA} not being able to match either the MaNGA results or the dwarf galaxies. However, neither of the two simulations managed to match the increase in $\log \mathrm{N}/\mathrm{O}$ values as a function of $\log \mathrm{O}/\mathrm{H} +12$ (which correlates to the increase in stellar mass), although the \texttt{SIMBA-C} simulation is showing a trend resembling the \citet{Dopita2016} observational average.

We also include the theoretical \texttt{Chem5} results from the \citet{Vincenzo2018b} study, which was done on a specific type of galaxy with specifications of $M_{*} = 3.62\times 10^{10}\,\textup{M}_\odot$ for the stellar mass, a gas fraction with respect to the total baryonic mass of $f_{gas} = 0.35$, an average stellar N/O ratio of $\log \mathrm{N}/\mathrm{O}=-0.85$, and lastly that the galaxy is forming stars. Furthermore, the results probe different spatial locations in this galaxy. The results for this galaxy are shown with the green region in Fig. \ref{fig: NO_vs_OH}. We therefore do not plot the exact same results, but nevertheless, the comparison allows insight into the \texttt{Chem5} model. Our three highest concentration levels all lie within this region, and our concentrated regions tend in the same direction, albeit with a larger spread. This larger spread in the results is likely due to including a larger stellar mass range. 

\citet{Vincenzo2018a} showed that their results do not match the slope from the observations and postulate that this might be a consequence of the lack of super-solar metallicity yields from AGB stars not being available\footnote{Nitrogen is mostly synthesised by intermediate-mass stars by hot bottom burning during the AGB phase, while oxygen is mostly produced by core-collapse SNe with short lifetimes ($\sim 10^{6} \mathrm{\, yr}$) \citep{Vincenzo2018a}.}; note that super-massive-AGB stars are included in both Chem5 and \citet{Vincenzo2018a, Vincenzo2018b}. The exact cause needs further investigation.

From the various chemical abundance tests, we can conclude that \texttt{SIMBA-C} shows several improvements over the \texttt{SIMBA} simulation's instantaneous recycling model and that it allows for more accurate chemical enrichment, abundance, and metallicity profile predictions in future work.

\section{Summary}\label{sec: Conclusions}
In this work we integrated the \texttt{Chem5} model, developed in \citet{Kobayashi2007}, into \texttt{SIMBA} \citep{Dave2019}. The \texttt{Chem5} model is a set of interpolation tables that determine the mass, energy, and yields from stellar populations based on stellar feedback events, such as stellar winds, AGB winds, SNe Ia, SNe II, and `failed' SNe. We also introduced the Chabrier IMF into the \texttt{Chem5} model to match the stellar populations between the \texttt{Chem5} model and the \texttt{SIMBA} simulation.

We adjusted our modified version of \texttt{SIMBA}, since we were underproducing metals at low redshifts, leading to a low metal cooling efficiency. Due to this, the stellar feedback strength was lowered. This led us to match the global stellar wind feedback velocity normalisation of \citet{Muratov2015}. We also improved the GSMF compared to the \texttt{SIMBA} simulation, especially at $z=0$. No direct tuning of the metallicities was performed. We only used its results to calibrate the simulation's SFR to the new model.

Other notable results include an improved sSFR--stellar mass relation for the low-and high-mass ranges while producing fewer galaxies in the green valley at intermediate masses, in better accordance with observational data. Furthermore, \texttt{SIMBA-C} shows improvements across all galaxy mass ranges in terms of the black hole--stellar mass relation relative to \texttt{SIMBA} in evolving black holes within low-mass galaxies, although this contributed to the new black hole seeding and growth implementation rather than the \texttt{Chem5} model. 

In terms of the metal content predicted by the new \texttt{Chem5} model, \texttt{SIMBA-C} was able to reproduce the observational data on almost all levels except for the stellar mass--metallicity relation at the highest galaxy masses. The new model will enable us to study individual chemical abundance more robustly.  These include an improved agreement in the [O/Fe] and [Mg/Fe] vs. [Fe/H] for stars within Milky Way-like galaxies and a better match to observations of log $\mathrm{N}/\mathrm{O}$ vs. $\log \mathrm{O}/\mathrm{H} +12$.

Future work will investigate how to extend the $\alpha$-abundance ratios to have a more distinct plateau for $[\mathrm{Fe}/\mathrm{H}]$$<-1.0$, as well as trying to constrain the large spread in the values of $\log \mathrm{N}/\mathrm{O}$ that occurs at low $\log \mathrm{O}/\mathrm{H} +12$. Finally, we will continue to make improvements to the \texttt{SIMBA-C} model such as including \texttt{SIMBA}'s dust model and explore how to alleviate \texttt{SIMBA-C}'s discrepancies with the gas-phase metal content of the high-mass galaxies.

We conclude that the integration of the new chemical enrichment model and other changes of the \texttt{SIMBA-C} model yields significant improvements over \texttt{SIMBA}. Moreover, the new model includes all elements from H $\rightarrow$ Ge, rather than just the standard 11 elements of stellar evolution. \texttt{Chem5} accounts for state-of-the-art enrichment processes such as `failed' SNe, super AGB, and HNe to generate these additional elements.  These advances open up new avenues for constraining chemical evolution processes within galaxies and circum-galactic and inter-galactic gas, which we will explore in future work.


\section*{Acknowledgements}

This work is based on research supported in part by the National Research Foundation of South Africa (NRF Grant Number: 146053). The analysis reported in this article was enabled by HPC resources provided by WestGrid and Digital Research Alliance of Canada (alliancecan.ca) award to AB. RH also acknowledges the \texttt{SIMBA} collaboration for the use of the simulation. RH also acknowledges the Royal Society of Science travel grant that allowed for in-person collaboration at the University of Edinburgh, Scotland, to further this study. AB and DR acknowledge support from the Natural Sciences and Engineering Research Council of Canada (NSERC) through its Discovery Grant program. Also, AB acknowledges support from the Infosys Foundation via an endowed Infosys Visiting Chair Professorship at the Indian Institute of Science. Additionally, AB acknowledges the l{\fontencoding{T4}\selectfont\M{e}}\'{k}$^{\rm w}${\fontencoding{T4}\selectfont\M{e}\m{n}\M{e}}n peoples on whose traditional territory the University of Victoria stands, and the Songhees, Equimalt and WS\'{A}NE\'{C} peoples whose historical relationships with the land continue to this day. Any opinion, finding, and conclusion or recommendation expressed in this material is that of the author(s), and the NRF does not accept any liability in this regard. WC is supported by the STFC AGP Grant ST/V000594/1, the Atracci\'{o}n de Talento Contract no. 2020-T1/TIC-19882 granted by the Comunidad de Madrid in Spain and the Project grant PID2021-122603NB-C21 from the Ministerio de Ciencia e Innovación, Spain. CK acknowledges funding from the UK Science and Technology Facility Council (STFC) through grant ST/R000905/1 and ST/V000632/1. The authors thank the anonymous reviewer for constructive comments that led to the improvement of the article.

\section*{Data Availability}
The published \texttt{SIMBA} simulation \citep{Dave2019} is available on the \texttt{SIMBA} university repository at \url{http://simba.roe.ac.uk/}. The \texttt{SIMBA-C} simulation data underlying this article will be shared on reasonable request to the corresponding author.



\bibliographystyle{mnras}
\bibliography{mnras_template} 


\bsp	
\label{lastpage}
\end{document}